\documentclass{emulateapj}
\usepackage{epstopdf}

\newcommand{\kms}{km~s$^{-1}$}

\newcommand{\ldl}{$\lambda/{\Delta}{\lambda}$}
\newcommand{\teff}{T$_{eff}$}
\newcommand{\logg}{$\log{g}$}
\newcommand{\fsed}{$f_{sed}$}
\newcommand{\vtan}{$V_{tan}$}

\newcommand{\lii}{\ion{Li}{1}}
\newcommand{\ki}{\ion{K}{1}}
\newcommand{\nai}{\ion{Na}{1}}
\newcommand{\csi}{\ion{Cs}{1}}
\newcommand{\rbi}{\ion{Rb}{1}}
\newcommand{\meth}{CH$_4$}
\newcommand{\water}{H$_2$O}
\newcommand{\wat}{H$_2$O}
\newcommand{\hh}{H$_2$}
\newcommand{\name}{2MASS~J11263991$-$5003550}
\newcommand{\namesh}{2MASS~J1126$-$5003}

\slugcomment{Accepted for publication to ApJ}

\shorttitle{The Blue L dwarf {\namesh}}
\shortauthors{Burgasser et al.}

\begin{document}

\title{Clouds, Gravity and Metallicity in Blue L dwarfs: The Case of {\name}\altaffilmark{8}}

\author{Adam J.\ Burgasser\altaffilmark{1,2},
Dagny L.\ Looper\altaffilmark{2,3}
J.\ Davy Kirkpatrick\altaffilmark{4}, 
Kelle L.\ Cruz\altaffilmark{5,6}, and
Brandon J.\ Swift\altaffilmark{3,7}}

\altaffiltext{1}{Massachusetts Institute of Technology, Kavli Institute for Astrophysics and Space Research,
Building 37, Room 664B, 77 Massachusetts Avenue, Cambridge, MA 02139; ajb@mit.edu}
\altaffiltext{2}{Visiting Astronomer at the Infrared Telescope Facility, which is operated by
the University of Hawaii under Cooperative Agreement NCC 5-538 with the National Aeronautics
and Space Administration, Office of Space Science, Planetary Astronomy Program.}
\altaffiltext{3}{Institute for Astronomy, University of Hawaii, 2680 Woodlawn Drive, Honolulu, HI 96822}
\altaffiltext{4}{Infrared Processing and Analysis Center, M/S
100-22, California Institute of Technology, Pasadena, CA 91125}
\altaffiltext{5}{Astronomy Department, California Institute of Technology, Pasadena, CA 91125}
\altaffiltext{6}{NSF Astronomy and Astrophysics Postdoctoral Fellow}
\altaffiltext{7}{Currently at Steward Observatory, University of Arizona, 933 N. Cherry Ave.,
Tucson, AZ 85721}

\altaffiltext{8}{This paper includes data gathered with the 6.5 meter Magellan Telescopes located at Las Campanas Observatory, Chile.}

\begin{abstract}
Optical and near-infrared spectroscopy of the newly discovered peculiar 
L dwarf {\name} are presented.
Folkes et al.\ identified this source as a high proper motion
L9$\pm$1 dwarf based on its  
strong H$_2$O absorption at 1.4~$\micron$.  
We find that the optical spectrum of {\namesh} is in fact consistent with that 
of a normal L4.5 dwarf with notably enhanced FeH absorption at 9896~{\AA}.
However, its near-infrared spectrum is unusually blue, with strong
H$_2$O and weak CO bands similar in character 
to several recently identified ``blue L dwarfs''.  Using 
{\namesh} as a case study, and guided by trends in the condensate
cloud models of Burrows et al.\ and Marley et al.,
we find that the observed spectral peculiarities of these sources 
can be adequately explained by the 
presence of thin and/or large-grained condensate clouds
as compared to normal field L dwarfs.
Atypical surface gravities or metallicities alone cannot reproduce
the observed peculiarities, although they may be partly responsible for the unusual
condensate properties.
We also rule out unresolved multiplicity as a cause for the
spectral peculiarities of {\namesh}.
Our analysis is supported by examination of {\em Spitzer}
mid-infrared spectral data from Cushing et al.\
which show that bluer L dwarfs tend to have weaker 10~$\micron$
absorption, a feature tentatively associated with silicate oxide 
grains.
With their unique spectral properties, 
blue L dwarfs like {\namesh} should prove useful in studying
the formation and properties of condensates and condensate clouds 
in low temperature atmospheres.
\end{abstract}

\keywords{stars: atmospheres ---
stars: fundamental parameters ---
stars: individual ({\name}) ---
stars: low mass, brown dwarfs
}

\section{Introduction}

L dwarfs comprise one of the two latest-type spectral
classes of very low mass stars and brown dwarfs,
spanning masses at and below the hydrogen burning minimum mass
(see \citealt{kir05} and references therein).
They are inexorably linked to the presence and properties of 
liquid and solid condensates
which form in their cool photospheres (e.g., \citealt{tsu96,bur99,ack01,all01,coo03,tsu05}).
These condensates 
significantly influence the
spectral energy distributions and photospheric gas abundances
of L dwarfs, by removing gaseous TiO and VO from the photosphere and 
enabling the retention of atomic alkali species
(e.g., \citealt{feg96,bur99,lod02}). Weakened {\wat} absorption through backwarming
effects (e.g., \citealt{jon97,all01}) and red 
near-infrared colors ($J-K$ $\approx$ 1.5--2.5; \citealt{kir00})
also result from condensate opacity.
In addition, periodic and aperiodic
photometric variability observed in several L dwarfs 
has been associated with surface patchiness in photospheric
condensate clouds (e.g., \citealt{bai99,bai01,gel02,moh02}).
Condensate abundances at the photosphere appear to reach their zenith 
amongst the mid- and late-type
L dwarfs \citep{kir99,cha00,ack01} before disappearing
from the photospheres of cooler T dwarfs 
\citep{mar96,tsu96b,all01,cus06}.
The abundances of photospheric condensates, their grain size
distribution, and the radial 
and surface structure of condensate clouds
may vary considerably from source to source, as 
well as temporally for any one source, and the dependencies
of these variations on various physical parameters are only
beginning to be explored \citep{hel01,woi03,kna04}. 

With hundreds of L dwarfs 
now known,\footnote{A current list is 
maintained at \url{http://dwarfarchives.org}.} groupings of
peculiar L dwarfs -- sources
whose spectral properties diverge consistently
from the majority of field objects --
are becoming distinguishable.
Examples include young, low surface gravity brown dwarfs
\citep{mcg04,kir06,all07,cru07} and metal-poor L subdwarfs
\citep{me0532,lep1610,giz06,megmos}.
There also exists a class of peculiar 
``blue'' L dwarfs
\citep{cru03,cru07,kna04}, roughly a dozen 
sources exhibiting normal optical spectra but unusually
blue near-infrared colors and strong near-infrared {\wat}, FeH 
and {\ki} features.  Various studies have attributed these peculiarities
to subsolar metallicity, high surface gravity, unresolved multiplicity and 
peculiar cloud properties 
\citep{giz00,cru03,cru07,mcl03,mewide3,kna04,chi06,fol07}.
Any one of these characteristics may impact the presence and
character of condensates and condensate 
clouds in low temperature atmospheres.

In an effort to identify new nearby and peculiar L dwarfs, 
we have been searching for
late-type dwarfs using near-infrared imaging
data from the Deep Near Infrared Survey of the Southern Sky 
(DENIS; \citealt{epc97}).  
One of the objects identified in this program is 
DENIS~J112639.9$-$500355, a bright source which 
was concurrently discovered
by \citet{fol07}
in the SuperCOSMOS Sky Survey 
\citep[hereafter SSS]{ham01a,ham01b,ham01c} and the
Two Micron All Sky Survey (hereafter 2MASS; \citealt{skr06}).
It is designated {\name} in that study, and 
we refer to the source hereafter as {\namesh}.
Based on its blue near-infrared colors and deep {\water} 
absorption bands, \citet{fol07} concluded that {\namesh} 
is a very late-type L dwarf (L9$\pm$1) which may have unusually
patchy or thin condensate clouds.  In this article, we critically examine
the observational properties of {\namesh} to unravel the origins of its spectral
peculiarities, and examine it as a representative of
the blue L dwarf subgroup.

Our identification of {\namesh} and a slightly revised determination
of its proper motion using astrometry from the SSS, DENIS and 2MASS catalogs
are described in $\S$~2.
Optical and near-infrared 
spectroscopic observations and their results are described in $\S$~3, along with  determination of the optical and near-infrared classifications
of {\namesh} and estimates of its distance
and space kinematics.  In $\S$~4 we analyze the properties of
{\namesh} and blue L dwarfs in general, considering metallicity,
surface gravity, condensate cloud and unresolved multiplicity
effects.  We also introduce a new near-infrared {\wat} index that
eliminates discrepancies between optical and near-infrared
types for these sources.
Results are discussed and summarized in $\S$~5.

\section{Identification of {\namesh}}

We initially identified {\namesh} in the DENIS
Data Release 3 point source catalog as part of a sample
constrained to have 9 $\leq J \leq$ 15.5, 
$I-J \ge 3$ (corresponding to spectral types M8 and later),
$J-K_s \leq 2.8$ (to exclude background giants),
galactic latitudes ${\mid}b{\mid} \geq$ 8$\degr$ (excluding
the Galactic plane) and declinations $-90\degr \leq \delta \leq +2\degr$.
Further details on this search sample and resulting 
discoveries will be made in a future publication (D.\ Looper et al., in preparation).
The combined DENIS and 2MASS colors of {\namesh} are 
$I-J$ = 3.80$\pm$0.15 and $J-K_s$ = 1.17$\pm$0.04,
consistent with a late type star or brown dwarf 
(e.g., \citealt{del97,kir00}).

\begin{figure}
\epsscale{1.0}
\plotone{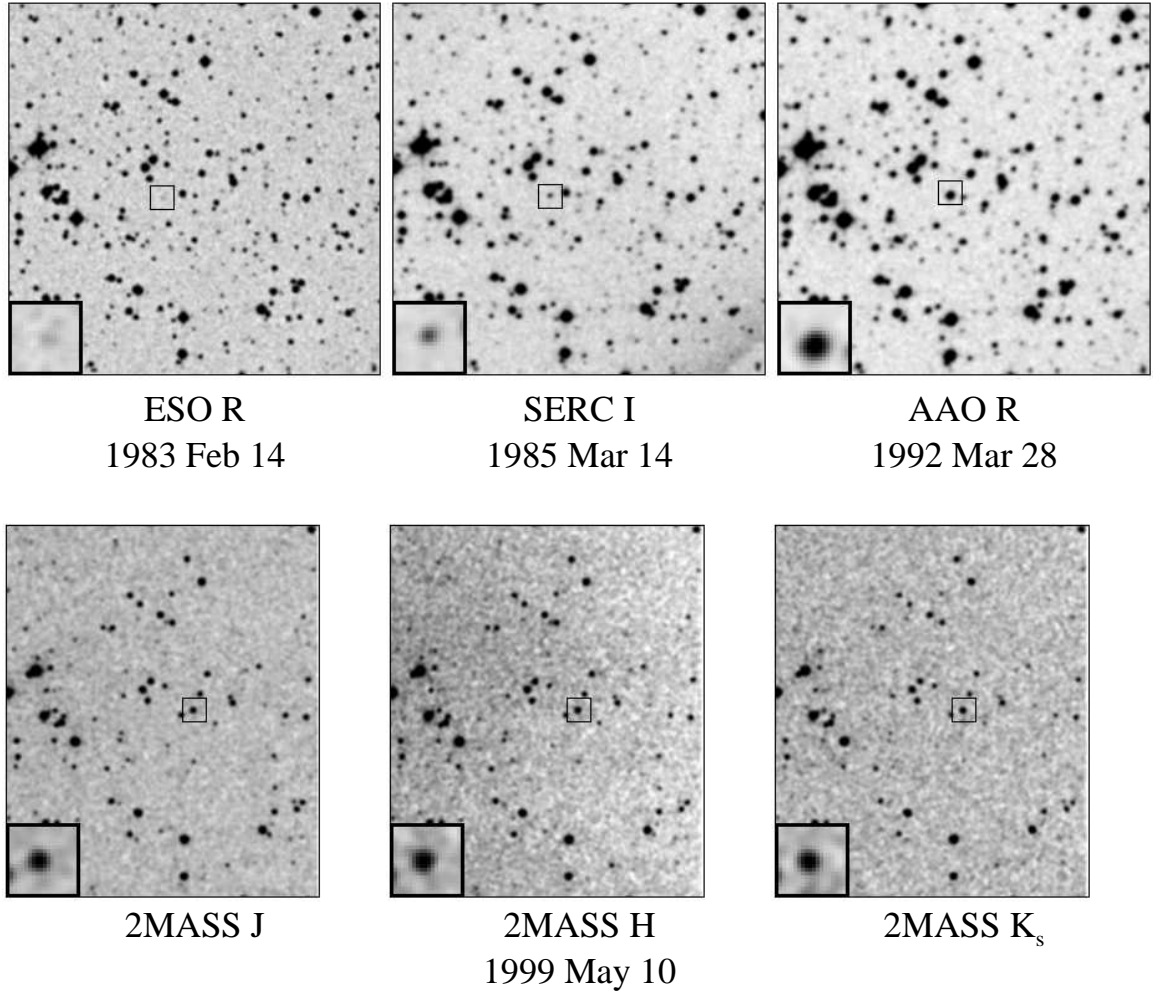}
\caption{Field images of {\name} from ESO $R$ (top left), 
SERC $I_N$ (top middle) and AAO $R$ (top right) photographic plates; and
2MASS $JHK_s$ (bottom).
All images are scaled to the same spatial
resolution and oriented with north up and east to the left.  Photographic plate images are 5$\arcmin$ on a side.  
Inset boxes 20$\arcsec$$\times$20$\arcsec$ in size in each image indicate the 
position of the source after correcting for its motion 
($\mu = 1{\farcs}66{\pm}0{\farcs}03$~yr$^{-1}$ at position angle
$\theta = 285{\fdg}3{\pm}1{\fdg}6$) and are expanded
in the lower left corner.
Note that the bright optical
source close to the motion-corrected position
of {\namesh} in the 1992 AAO $R$ image is a background star.
\label{fig_finder}}
\end{figure}

Figure~\ref{fig_finder} displays optical field images centered on
the 2MASS/DENIS coordinates of {\namesh} as observed by the ESO $R$
(epoch 1983 February 14 UT), SERC $I_N$
(epoch 1985 March 14 UT) and AAO $R$ (epoch 1992 March 28 UT)
photographic plate surveys (e.g., \citealt{har81,can84}).  
There are no optical counterparts to {\namesh} 
in these images at the exact 2MASS/DENIS position, but
faint, offset counterparts can be discerned.
Based on SSS and 2MASS astrometry spanning 14.2~yr, 
\citet{fol07} determined a substantial
proper motion for {\namesh}, 
$\mu = 1{\farcs}65{\pm}0{\farcs}03$~yr$^{-1}$.
We confirm that the faint 
$R$-band (20.4~mag) and $I_N$-band (17.6~mag\footnote{Optical photometry as given in the SSS.}) counterparts
in the 1983 and 1985 photographic
plate images are at the expected position of {\namesh} 
based on this motion, neither of which have coincident
near-infrared counterparts.\footnote{A brighter optical
star is coincident with the motion-corrected position
of {\namesh} in the 1992 AAO $R$ image and obscures the 
proper motion source.}
The associated optical/near-infrared colors
($R_{ESO}-J$ = 6.4; $I_N-J$ = 3.6) are again indicative of a late-type
dwarf. A linear fit to the
SSS, DENIS and 2MASS astrometry over 16.3~yr (Table~\ref{tab_astrometry})
yields a value
of $\mu = 1{\farcs}66{\pm}0{\farcs}03$~yr$^{-1}$ at position angle
$\theta = 285{\fdg}3{\pm}1{\fdg}6$, where uncertainties
include an estimated 0$\farcs$3 astrometric uncertainty in both Right Ascension
and declination for all three catalogs.  Not surprisingly, this 
value is consistent with the measurement of \citet{fol07}. 
Note that neither proper motion measurement takes
into account parallactic reflex
motion, which is presumably much smaller than the 
aggregate linear motion of {\namesh} since 1983 
(nearly 30$\arcsec$).

\section{Spectroscopic Observations}

\subsection{Optical Data}

Optical spectroscopy of {\namesh} was obtained
on 2006 May 8 (UT) using the Low Dispersion Survey Spectrograph (LDSS-3)
mounted on the Magellan 6.5m Clay Telescope.
LDSS-3 is an imaging spectrograph, upgraded from the original
LDSS-2 \citep{all94} for improved
red sensitivity.   Conditions during the observations were
clear with moderate seeing (0$\farcs$7 at $R$-band).
We employed the VPH-red grism (660 lines/mm) with a 0$\farcs$75
(4 pixels) wide longslit mask, aligned to the parallactic angle, to
acquire 6050--10500 {\AA} spectra across the entire chip
with an average resolution of {\ldl} $\approx$ 1800.  Dispersion
along the chip was 1.2~{\AA}/pixel.  The OG590 longpass filter
was used to eliminate second order light shortward of 6000~{\AA}.
A single slow-read exposure of 750~s was obtained at an airmass
of 1.08. We also observed the
G2~V star HD~97625 immediately after the {\namesh} observation and
at a similar airmass
for telluric absorption correction.  The flux standard LTT 7987 
(a.k.a.\ GJ 2147; \citealt{ham94}) was observed 
during the same run on 2006 May 7 (UT) using an identical slit 
and grism combination. All spectral observations 
were accompanied by HeNeAr arc lamp and flat-field
quartz lamp exposures for dispersion and pixel response calibration.

LDSS-3 data were reduced in the IRAF\footnote{IRAF is 
distributed by the National Optical
Astronomy Observatories, which are operated by the Association of
Universities for Research in Astronomy, Inc., under cooperative
agreement with the National Science Foundation.} environment.
Raw science images (separated into short and long wavelength halves)
were first trimmed and subtracted by a median combined
set of slow-read bias frames taken during the afternoon.  
The resulting images were then divided by the corresponding
normalized, median-combined and bias-subtracted set of flat field frames.
The G star spectra were first
extracted using the APALL task, utilizing background subtraction
and optimal extraction options.  The spectrum of {\namesh}
was extracted using
the G star dispersion trace as a template.  Dispersion solutions were
determined from the arc lamp spectra extracted 
using the same dispersion trace; 
solutions were accurate
to $\sim$0.1 pixels, or $\sim$0.12~{\AA}.  
Flux calibration was determined
using the tasks STANDARD and SENSFUNC with
observations of LTT 7987, adequate over the
spectral range 6000--10000 {\AA}. Corrections to telluric O$_2$ (6860--6960 {\AA} B-band,
7580--7700 {\AA} A-band)
and H$_2$O (7150--7300 {\AA}) absorption bands
were determined by linearly interpolating over these features in the 
G dwarf spectrum, dividing by the uncorrected spectrum, 
and multiplying the
result with the spectrum of {\namesh}.  Note that we did not correct for
9270--9675 {\AA} telluric H$_2$O absorption due to 
the reduced signal at these wavelengths in the target and G dwarf
spectra.
Short and long wavelength data were then
stitched together with no additional flux scaling.

\begin{figure}
\epsscale{0.7}
\plotone{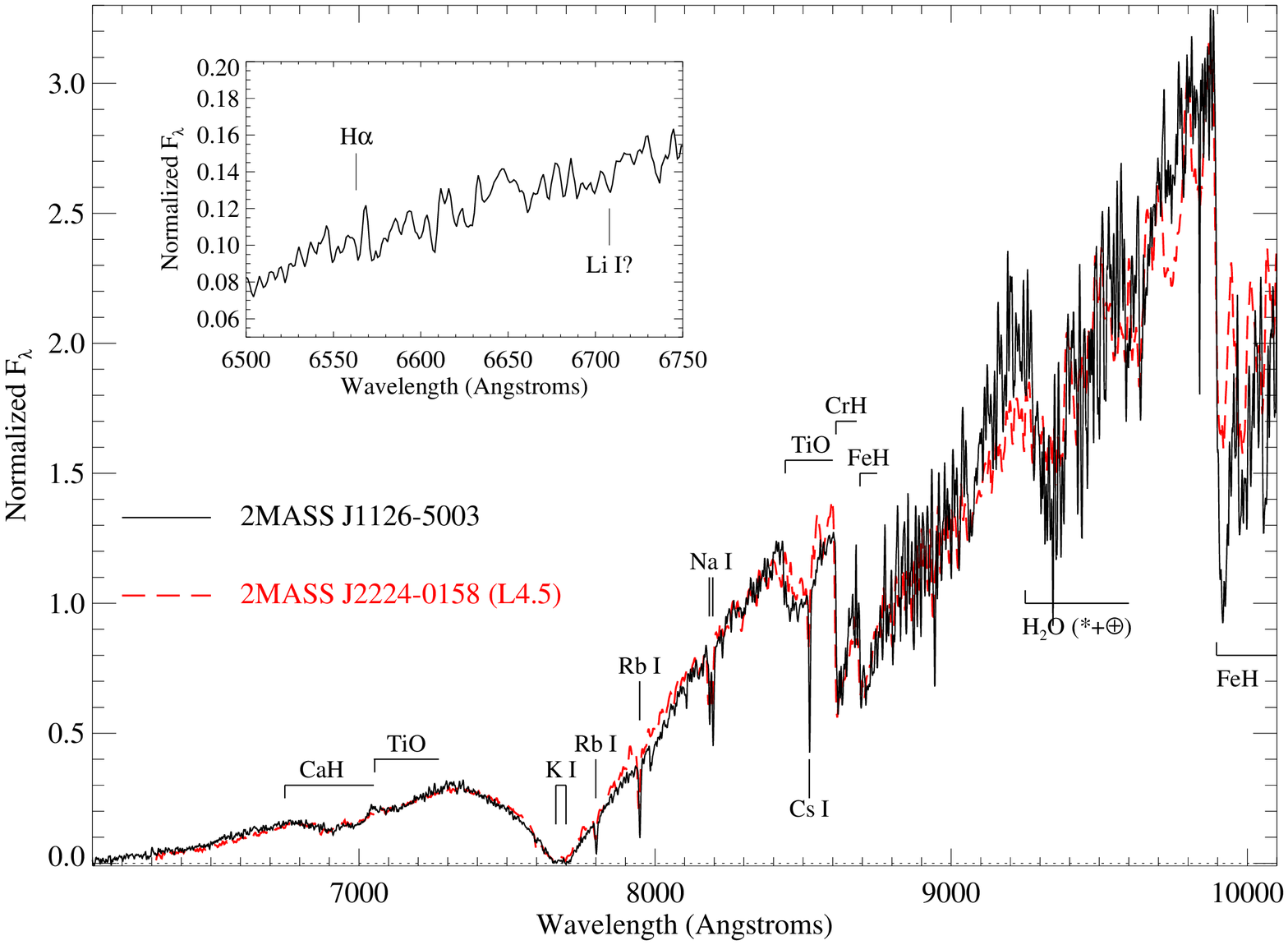}
\plotone{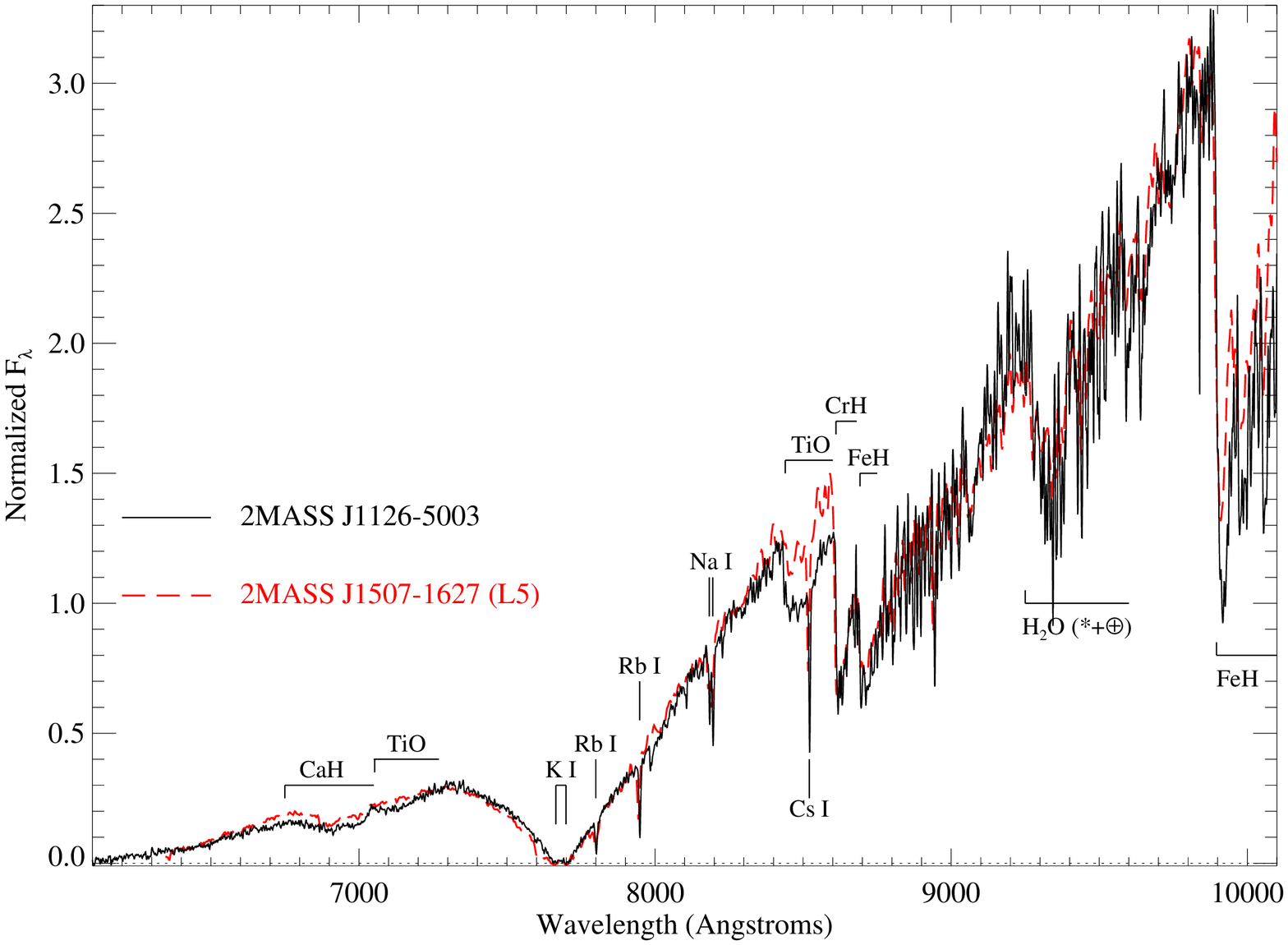}
\caption{Red optical (6100--10100~{\AA}) spectrum of {\namesh} (black lines)
obtained with LDSS-3, compared to the LRIS spectra of the
L4.5 2MASS~J2224-0158 (top panel, red dashed line)  
and the L5 2MASS~J1507-1627 (bottom panel, red dashed line) from \citet{kir00}.  
All spectra are normalized at 8200~{\AA}. 
Note that the spectrum of {\namesh}
has been corrected for telluric absorption shortward of 9000~{\AA},
while those of 2MASS~J2224-0158 and 2MASS~J1507-1627
have not. 
Major molecular and atomic
absorption features are labelled.  The inset box in the top panel
expands the 6500--6750~{\AA} spectrum of {\namesh} hosting the 6563~{\AA} H$\alpha$ emission and
6708~{\AA} \ion{Li}{1} absorption lines, neither of which are 
convincingly detected.
\label{fig_optspec}}
\end{figure}

\begin{figure}
\epsscale{1.0}
\plottwo{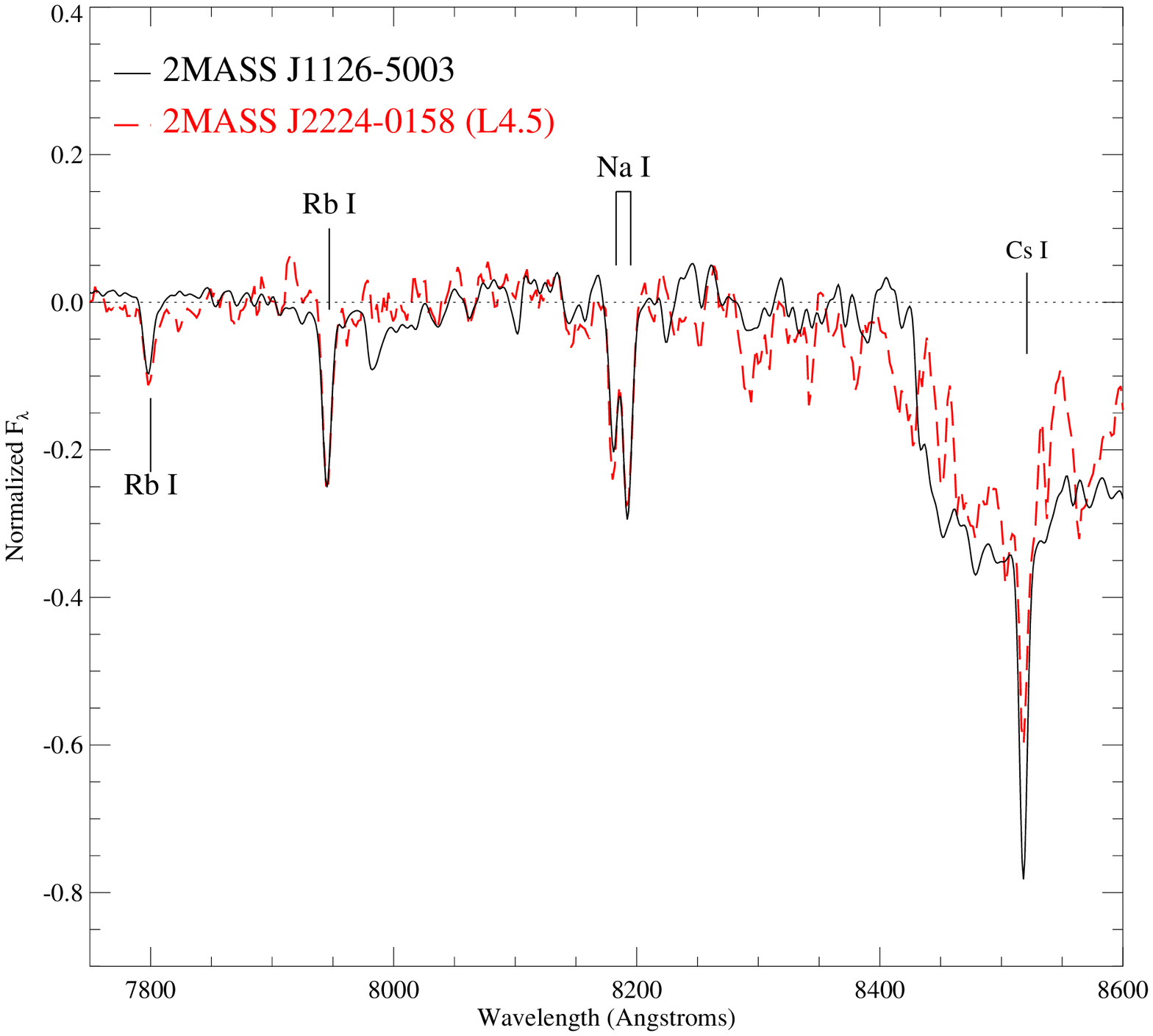}{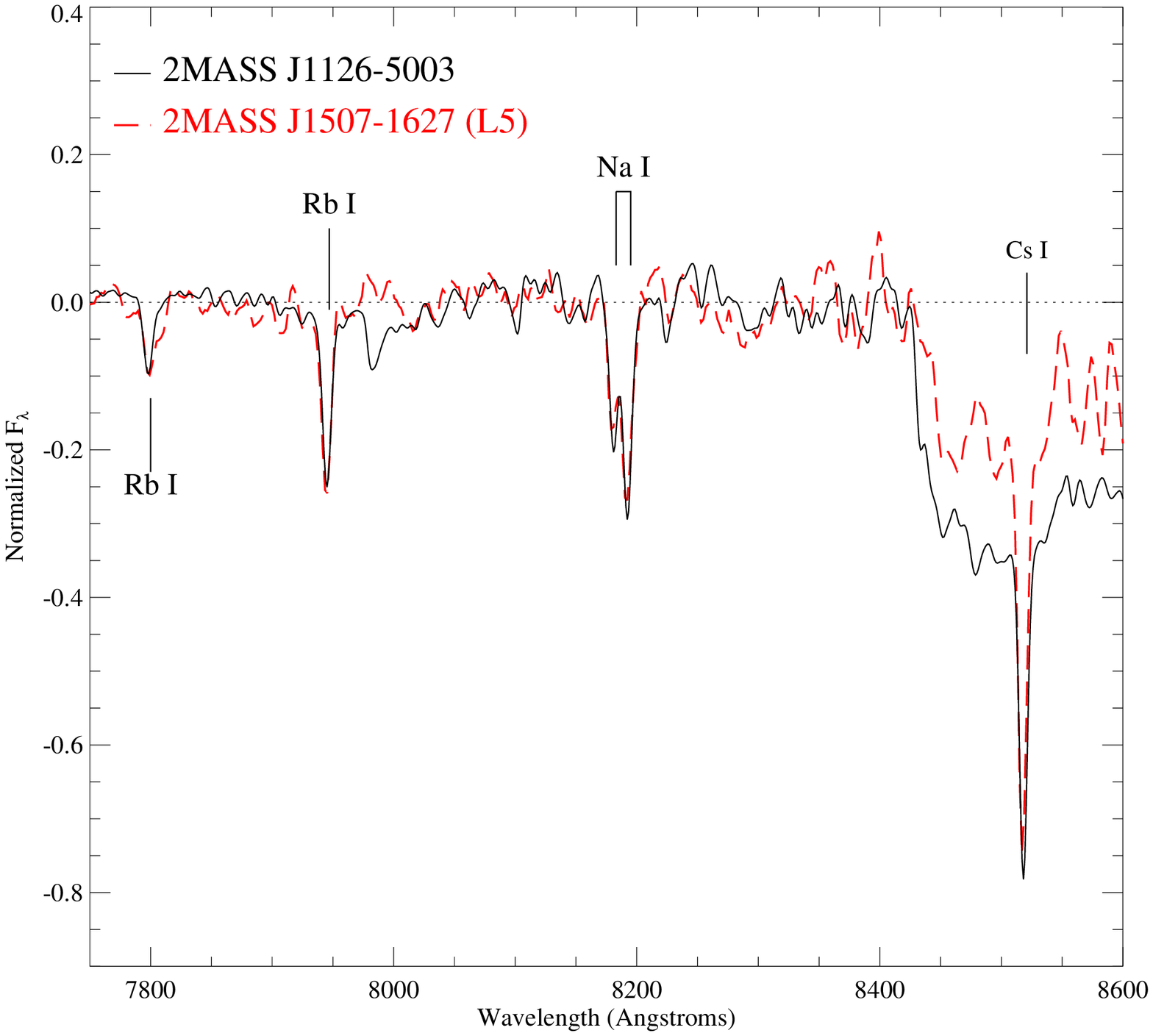}
\caption{Red optical alkali lines ({\rbi}, {\nai} and {\csi})
in the 7750--8300~{\AA} spectal region of {\namesh} (black lines)
compared to the L4.5 2MASS~J2224-0158 (left panel, red dashed line)
and the L5 2MASS~J1507-1627 (right panel, red dashed line).
A linear fit to the local continuum in this spectral range has
been subtracted from all spectra to highlight the line absorption, and
data have been deconvolved to the same resolution ({\ldl} = 1000)
and shifted to their frame of rest for accurate comparison.
\label{fig_alkalines}}
\end{figure}

The reduced red optical spectrum of {\namesh}
is shown in Figure~\ref{fig_optspec}, compared to the
L4.5 2MASS J22244381-0158521 \citep[hereafter 2MASS~J2224-0158]{kir00}
and the L5 2MASS~J1507476-162738 
\citep[hereafter 2MASS~J1507-1627]{rei00}\footnote{The optical
spectra of 2MASS~J2224-0158 and 2MASS~J1507-1627
were obtained using the Low Resolution
Imaging Spectrograph (LRIS, \citealt{oke95}) 
mounted on the Keck 10m Telescope,
and reduced by J.\ D.\ Kirkpatrick using identical IRAF routines. 
Comparisons of late-type spectra observed with both instruments
show consistency to within 10\% over the 6100--9000~{\AA}
range \citep{megmos}.}. 
The overall optical spectral morphology of {\namesh} is well-matched
to both L dwarf comparison sources, agreeing best with 
the L4.5 shortward of the pressure-broadened 
7665/7699 {\AA} {\ki} doublet \citep{bur00,all03,bur03}
and the L5 longward of this feature.
{\namesh} exhibits the same peak-up in flux 
between the blue wing of {\ki} and the red wing of
the pressure-broadened 5890/5896 {\AA} {\nai} D lines 
present in the L dwarf comparison spectra,
as well as line absorption from 
{\rbi} (7800 and 7948 {\AA}),
{\nai} (8183/8195~{\AA} doublet)
and {\csi} (8521~{\AA}).  These
lines are shown in detail in Figure~\ref{fig_alkalines}, and
their equivalent widths (EW) are listed in Table~\ref{tab_ews}.
Line strengths are similar to those of 
other midtype field L dwarfs \citep{kir99}.
The optical spectrum of {\namesh} also exhibits
strong metal hydride bands at 6950~{\AA} (CaH), 8600~{\AA} (CrH and FeH)
and 9896~{\AA} (FeH); and weak TiO absorption at 7100 and 8400~{\AA}.
The 9896~{\AA} Wing-Ford band of FeH is clearly stronger
in the spectrum of {\namesh} as compared to either of the L dwarf
comparison sources, while the 8400~{\AA} TiO is also slightly deeper,
particularly in contrast
to 2MASS~J1507-1627.  These features suggest that {\namesh} could be slightly
metal-poor, exhibiting the same peculiarities as L subdwarfs \citep{me0532,megmos}, a point that is discussed further
in $\S$~4.2.2. 

A close examination at the 6500--6750~{\AA} region 
(inset in Figure~\ref{fig_optspec}) reveals no significant
emission from the 6563~{\AA} H$\alpha$ line, 
an indicator of magnetic activity.  The absence of H$\alpha$ 
is consistent with the general decline
in optical magnetic emission between late-type M to midtype L dwarfs 
\citep{giz00,moh03,wes04}.  There is a weak feature at the
location of the 6708~{\AA} {\lii} line,
an indicator of substellar mass \citep{reb92}, but is of the same 
strength as noise features in this spectral region.
The upper limit EW of the {\lii} line ($<$ 0.4~{\AA})
is considerably less than the minumum measured EWs of detected lines 
in other L-type brown dwarfs ($\sim$3~{\AA}; \citealt{kir00}).
We therefore conclude that {\lii} absorption is not present in the spectrum
of {\namesh}, although higher resolution, higher signal-to-noise observations
are necessary to confirm this result.

\subsection{Near-Infrared Data}

Low resolution near-infrared
spectral data for {\namesh} were
obtained in clear conditions
on 2006 December 20 (UT) using the SpeX spectrograph \citep{ray03}
mounted on the 3m NASA Infrared Telescope Facility (IRTF).
The 0$\farcs$5 slit was employed, providing 0.75--2.5~$\micron$
spectroscopy with resolution {\ldl} $\approx 120$
and dispersion across the chip of 20--30~{\AA}~pixel$^{-1}$.
To mitigate the effects of differential refraction, the slit was aligned
to the parallactic angle. Due to the southern declination
of this source, observations were made close to transit but at fairly 
high airmass (3.0).  12 exposures of 
60~s each were obtained 
in an ABBA dither pattern along the slit.
The A0~V star HD~101802 was observed immediately
afterward at a similar airmass (3.1) for flux calibration.
Internal flat field and Ar arc lamps were also observed
for pixel response and wavelength calibration.

Data were reduced using the SpeXtool package, version 3.3
\citep{cus04}, using standard settings.
Raw science images were first
corrected for linearity, pair-wise subtracted, and divided by the
corresponding median-combined flat field image.  Spectra were optimally extracted using the
default settings for aperture and background source regions, and wavelength calibration
was determined from arc lamp and sky emission lines.  The multiple
spectral observations were then median-combined after scaling individual
spectra to match the highest signal-to-noise
observation.  Telluric and instrumental response corrections for the science data were determined
using the method outlined
in \citet{vac03}, with line shape kernels derived from the arc lines. 
Adjustments were made to the telluric spectra to compensate
for differing \ion{H}{1} line strengths in the observed A0~V spectrum
and pseudo-velocity shifts.
Final calibration was made by
multiplying the spectrum of {\namesh} by the telluric correction spectrum,
which includes instrumental response correction through the ratio of the observed A0~V spectrum
to a scaled, shifted and deconvolved Kurucz\footnote{See \url{http://kurucz.harvard.edu/stars.html}.}
model spectrum of Vega. 

\begin{figure}
\epsscale{0.6}
\plotone{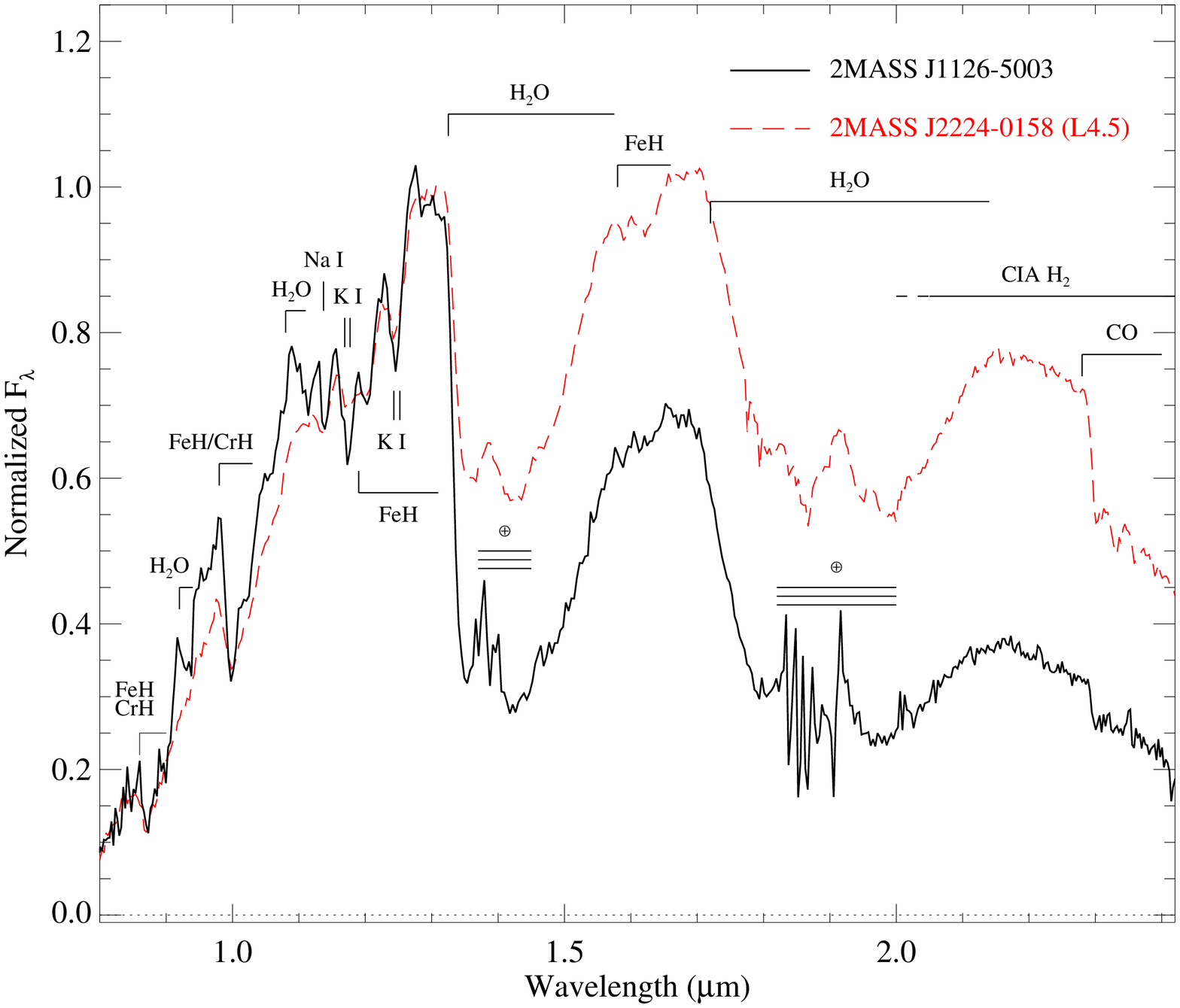}
\plotone{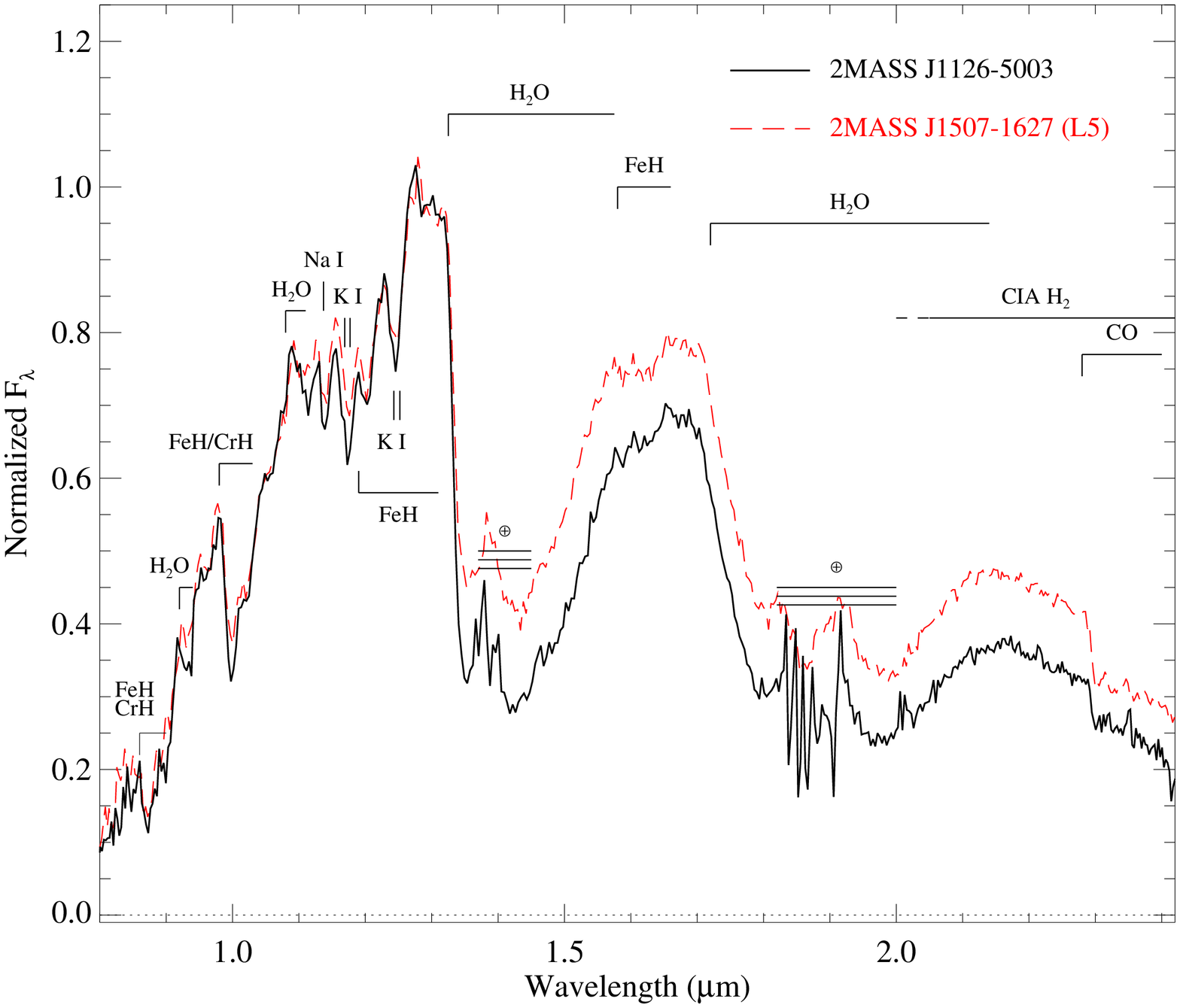}
\caption{Near-infrared SpeX prism spectra of {\namesh} (black line) 
compared to the L4.5 
2MASS~J2224-0158 (top panel, red dashed line).
and the L5
2MASS~J1507-1627 (bottom panel, red dashed line). 
All spectra
are normalized at 1.28~$\micron$.  Major molecular 
(FeH, CrH, {\water}, CO, H$_2$) and atomic
(\ion{Na}{1} and \ion{K}{1}) absorption features are labelled,
as well as regions of strong telluric absorption ($\oplus$).
\label{fig_nirspec}}
\end{figure}

The reduced spectrum of {\namesh} is shown in Figure~\ref{fig_nirspec} and
compared to equivalent SpeX prism data for 
2MASS~J2224-0158 (K.\ Cruz et al., in preparation)
and 2MASS~J1507-1627 \citep{me0805}.
While all three spectra exhibit features in common, 
including strong H$_2$O (1.4 and 1.9~$\micron$), 
CO (2.3~$\micron$) and FeH bands (0.99, 1.2 and 1.6~$\micron$),
and line absorption from {\ki} and {\nai} in the 1.1--1.25~$\micron$ region,
overall spectral morphologies differ markedly.
The near-infrared spectrum of {\namesh} is a better match to that of
2MASS~J1507-1627, but
is clearly bluer than both L dwarf comparison sources. 
This is consistent with its bluer near-infrared
colors, $J-K_s$ = 1.17$\pm$0.04 versus 2.05$\pm$0.04 
and 1.52$\pm$0.04 for 2MASS~J2224-0158 and
2MASS~J1507-1627, respectively.\footnote{We confirmed
that the flux calibration of the spectral data for both sources
was accurate by computing synthetic colors using 2MASS 
$J$ and $K_s$ relative spectral response curves from \citet{coh03}.  The synthetic colors agreed with photometric measurements to within 
their uncertainties.}  {\namesh} also exhibits
stronger {\wat} absorption 
and weaker FeH and CO absorption longward of 1.4~$\micron$,
although the strong 0.99~$\micron$ FeH band is again evident.
The deep {\wat} band at 1.4~$\micron$
was explicitly noted by \citet{fol07} and cited as evidence that this
source is a very late-type L dwarf.  
Indeed, we confirm that only L9 to T1
dwarfs have comparably strong {\wat} absorption (see Figure 2 in \citealt{fol07}), although the
absence of 1.6 and 2.2~$\micron$ {\meth} absorption bands
implies that {\namesh} is not a T dwarf \citep{meclass,geb02}.
\citet{fol07} also note relatively strong alkali line absorption in 
the 1.1--1.3 $\micron$ spectrum of {\namesh}, 
unresolved in our SpeX prism data.

Differences in the near-infrared spectral morphologies
of L dwarfs with similar
optical classifications but different $J-K_s$ colors
has been previously noted and discussed
in the literature (e.g., Fig.~24 in \citealt{mcl03}
and Fig.~8 in \citealt{me1520}).  
The spectral comparisons in Figures~\ref{fig_optspec} and~\ref{fig_nirspec}
serve to emphasize that these differences
are largely restricted to near-infrared wavelengths,
involving not just shifts in spectral color but in 
specific features as well.  The underlying physical
causes for these differences, particularly for blue
L dwarfs, are discussed further in $\S$~4.2.

\subsection{Spectral Classification}

Comparison of the optical spectrum of {\namesh} to those of 
2MASS~J2224-0158 and 2MASS~J1507-1627 in Figures~\ref{fig_optspec} indicate
a midtype L dwarf optical classification.
A more quantitative determination can be made by measuring 
the spectral ratios defined by
\citet{kir99}.  These values are 
listed in Table~\ref{tab_indices},\footnote{Note that spectral ratio measurements were
made after shifting the spectrum to its frame of rest;
see $\S$~3.4.} and yield an average subtype of L4.5$\pm$0.5.
Spectral ratios
from \citet{mrt99} and \citet{haw02} were also examined 
and yield consistent classifications
of L4--L5 on the \citet{kir99} scheme.
The consistency of these various indices, 
and the overall agreement between the spectra of {\namesh}
and 2MASS~J2224-0158 as shown in Figure~\ref{fig_optspec}, 
indicate a robust optical type of L4.5 for this source.

This classification disagrees significantly with the near-infrared
type determined by \citet{fol07}, L9$\pm$1, which is based largely on the
strength of the deep 1.4~$\micron$ {\wat} band.
This study also noted a ``duality'' in the near-infrared
characteristics of {\namesh}, with FeH features at 0.99 and 1.2~$\micron$
consistent with an early- to midtype L dwarf (based on indices
defined by \citealt{mcl03}).
As the \citet{fol07} spectrum spanned only the 1.0--1.6~$\micron$ region,
we re-examined the near-infrared type for this source using our broadband
1.0--2.5~$\micron$ SpeX data.

There is as yet no formal spectral classification
scheme for L dwarfs in the near-infrared; however, several 
studies have developed spectral index relations linked
to optical classifications.  
We examined spectral ratios for low-resolution near-infrared
data defined by \citet{tok99,rei01,geb02,meclass2}; and \citet{all07}, 
which sample the prominent {\wat} bands and details within the
spectral flux peaks.  
Values and associated spectral
types (based on polynomial relations determined in the studies listed above)
are listed in Table~\ref{tab_indices}. 
We derive an overall near-infrared spectral type of L6.5$\pm$2
based on the spectral type/index relations of
\citet{rei01}, with the uncertainty indicating the scatter in the index
subtypes.  This classification, while formally consistent
with that of \citet{fol07}, is clearly poorly constrained.
If only the indices sampling
the 1.4~$\micron$ {\wat} band are considered, an average type
of L8$\pm$1 is derived, in closer agreement with the result of  \citet{fol07}.  
However, indices sampling features at wavelengths
longward of 1.6~$\micron$ (i.e., K1, {\meth}~2.2~$\micron$ and {\meth}-K)
yield a mean type of L5$\pm$0.5, consistent with the optical classification
and in sharp disagreement with the {\wat} indices.  
The large discrepancy amongst the index subtypes simply reflects the fact that
none of the L dwarf spectral standards provide a good 
match to the near-infrared spectral energy distribution
of {\namesh}. In other words, its near-infrared spectrum
is truly peculiar.

\subsection{Estimated Distance and Kinematics}

Given its apparently robust 
optical spectral type, we chose to estimate the properties of {\namesh} by comparing
it to other optically-classified midtype L dwarfs.
A spectrophotometric distance estimate was determined
by comparing this source's 2MASS $JHK_s$ photometry
to seven absolute magnitude/spectral type relations from 
\citet{dah02,cru03,tin03}; and \citet{vrb04}.  The average distance derived
was 15$\pm$2~pc, which includes a $\pm$0.5 subclass 
uncertainty in the optical classification.  Distance estimates from
$J$-band photometry (14~pc) were slightly less than those from $K_s$-band
photometry (17~pc), consistent with the blue near-infrared colors of
this source relative to other L4--L5 dwarfs.  
Our estimated distance for {\namesh}
is nearly twice that of \citet{fol07}
based on their L9 near-infrared type.  Given the
better agreement in optical spectral morphology between {\namesh}
and other L4--L5 dwarfs, we contend that our larger distance
estimate is likely to be more accurate, assuming that {\namesh} is single 
(see $\S$~4.2.1).

The estimated distance and measured proper motion for {\namesh} implies 
a substantial tangential velocity, $V_{tan}$ = 117$\pm$15~{\kms}.  This is 
one of the highest $V_{tan}$s estimated or measured 
for any field L dwarf,\footnote{The L subdwarfs 
2MASS~J05325346+8246465 and 2MASS~J16262034+3925190
have considerably larger $V_{tan}$s consistent with their
halo kinematics \citep{me0532,me1626}.} surpassed only by 
the blue L3 dwarf 2MASSI~J1721039+334415 
\citep[hereafter 2MASS~J1721+3344]{cru03} with $V_{tan}$ = 139$\pm$15~{\kms} \citep{sch07}.
Indeed, only five out of $\sim$150 field L dwarfs 
with $V_{tan}$ determinations have values greater
than 100~{\kms}, including {\namesh} \citep[and references therein]{sch07}.

The radial velocity ($V_{rad}$) of {\namesh} was measured 
using the {\nai}, {\rbi} and {\csi} lines present in the
7800--8600~{\AA} region (Figure~\ref{fig_alkalines}).
Line centers were determined from Gaussian fits to the line cores and
compared 
to vacuum wavelengths listed in the Kurucz Atomic Line Database\footnote{Obtained
through the online database search form created by C.\ Heise and maintained
by P.\ Smith; see 
\url{http://cfa-www.harvard.edu/amdata/ampdata/kurucz23/sekur.html}.} 
\citep{kur95}.
The mean and standard deviation of velocity shifts for these
five lines gives $V_{rad}$ = 46$\pm$9~{\kms},
which includes a 5~{\kms} uncertainty in the dispersion solution of the optical
data.  The corresponding [$U,V,W$] space 
velocities of {\namesh} in the Local Standard of Rest (LSR),
assuming an LSR solar motion 
of [$U_{\sun},V_{\sun},W_{\sun}$] = [10,5,7]~{\kms} \citep{deh98},
is estimated as [$85,-98,-6$]~{\kms}.  These values are just within 
the 3$\sigma$
velocity dispersion sphere of local disk M dwarfs 
([$\sigma_U$,$\sigma_V$,$\sigma_W$] $\approx$ [40,28,19]~{\kms}
centered at [-13,-23,-7]~{\kms}; \citealt{haw96}), indicating 
that {\namesh} is likely to be an old disk or thick disk star. 
{\namesh} would appear to be
considerably older than
the average field L dwarf.  This is consistent
with the absence of \ion{Li}{1} absorption in its optical spectrum,
implying an age $\gtrsim$2~Gyr for a mass 
$>$0.065~M$_{\sun}$, assuming a best guess {\teff} $\approx$ 1700~K (typical
for L4--L5 dwarfs; \citealt{gol04,vrb04}) and solar metallicity
evolutionary models \citep{bur97}.   Table~\ref{tab_properties}
summarizes the estimated physical properties of
{\namesh}.  

\section{Analysis}

\subsection{{\namesh} in Context: The Blue L Dwarfs}

The discrepancies between the optical and near-infrared spectral
classifications of {\namesh}, and the near-infrared
spectral peculiarities noted in $\S$~3.2, are consistent
with the characteristics of blue L dwarfs reported in the literature.
This is illustrated in Figure~\ref{fig_blue}, which
compares the optical and near-infrared spectra of {\namesh} and 
three early-type blue L dwarfs ---
the L1 2MASS~J13004255+1912354 (hereafter 2MASS~J1300+1912; \citealt{giz00})
the L2 SIPS~J0921-2104 \citep{dea05}, and the L3 2MASS~1721+3344 ---
to ``normal'' L dwarfs with equivalent optical
classifications.\footnote{Additional optical spectral data shown here
are from \citet{kir00,cru03,cru07} and K.\ Cruz et al.\ (in preparation).
Additional SpeX prism spectral data shown are from 
\citet{me1520,me0805} and 
K.\ Cruz et al.\ (in preparation).}
All four sources are $\sim$0.3--0.5~mag
bluer than the average for their optical 
spectral type (e.g., \citealt{kir00}), 
and all show enhanced 1.4~$\micron$ {\wat}
absorption, weak CO absorption and unusually blue spectral 
energy distributions as compared to their normal L dwarf counterparts.  

In addition, there is consistent disagreement between optical and 
near-infrared classifications amongst these sources.  Using the 
spectral index relations
of \citet{rei01}, we find that near-infrared types are 
$\sim$2 subtypes later than optical types, 
larger than the uncertainties in these
relations; and the near-infrared indices themselves exhibit significant scatter.
The similarities between these L dwarfs suggests that their
spectral peculiarities have a common origin.

\subsection{Why are Blue L Dwarfs Peculiar?}

What drives these spectral peculiarities?
As noted in $\S$~1, various studies
have evoked unresolved multiplicity, subsolar metallicities,
high surface gravities and thin condensate clouds
as possible causes.  We examine each of these possibilities below,
focusing primarily on the properties of {\namesh}.

\subsubsection{Unresolved Multiplicity?}

Peculiar spectra commonly arise from the combined light of two
blended sources with differing spectral types.  Examples
include M dwarf plus white dwarf binaries (e.g., \citealt{wac03})
and L dwarf plus T dwarf 
binaries (e.g., \citealt{cru04,loo07}).
\citet{fol07} explicitly
considered this possibility for {\namesh} in their analysis.  Indeed, 
a case for unresolved multiplicity
can be made based on the apparent similarities of this 
source to the blue L dwarf 
SDSS~J080531.84+481233.0 (hereafter SDSS~J0805+4812; \citealt{haw02}),
which itself appears to be a binary \citep{me0805}.  
While \citet{fol07} reject unresolved multiplicity
as an explanation for the peculiarity of {\namesh}, we examine
this possibility again using our more comprehensive spectral coverage.

We compared the near-infrared spectrum
of {\namesh} to synthesized binary spectra constructed from 
SpeX prism data for a large sample of unresolved 
(i.e., apparently single) L and T dwarfs.
Our analysis was identical to that described
in \citet{me0805}, with binary spectra constructed by scaling the spectral templates
according to the $M_K$/spectral type relation of \citet{meltbinary}.
Figure~\ref{fig_double} displays the four best 
binary fits based on the  
minimum $\chi^2$ deviation between the normalized spectra.\footnote{Here, 
$\chi^2 \equiv \sum_{\{ \lambda\} }\frac{[f_{\lambda}(D1126)-f_{\lambda}(SB)]^2}{f_{\lambda}(D1126)}$, where $f_{\lambda}(D1126)$ is the spectrum of {\namesh}
and $f_{\lambda}(SB)$ the spectrum of the synthesized binary over the set
of wavelengths $\{ \lambda \}$ = 1.0--1.35, 1.45--1.8 and 2.0--2.35~$\micron$.
See \citet{me0805}.}
The best fitting pair, composed of 
the L3.5 2MASSW~J0036159+182110 \citep[hereafter 2MASS~J0036+1821]{rei00}
and the T4 2MASS~J21513839-4853542 \citep{ell05}, is a fairly good match
in the near-infrared,
particularly for the deep {\wat} bands and blue spectral 
energy distibution of {\namesh}.
However, the weak 1.6~$\micron$ {\meth} feature present in the
synthesized binary spectrum (also seen in the spectrum 
of SDSS~J0805+4812) is not present in the spectrum of {\namesh}.
Furthermore, this combination does not reproduce the optical spectrum
of {\namesh}, as illustrated in Figure~\ref{fig_doubleopt}.\footnote{In this Figure, the
spectrum of {\namesh} is compared to a binary template 
constructed from data for
2MASS~J0036+1821 \citet{rei00} and the T4.5 2MASS~J05591914-1404488 
\citep{me0559,me03opt}, as no optical T4 spectrum was available.  The component
spectra were scaled to the measured $M_{I_c}$ magnitudes of these sources
(16.41$\pm$0.02 and 19.11$\pm$0.07, respectively; \citealt{dah02}).}
In this case, the T dwarf secondary contributes negligibly to the optical flux of the binary,
and as a result the hybrid spectrum is nearly identical to that
of 2MASS~J0036+1821 (with the notable exception of weaker
TiO absorption at 8400~{\AA}) and inconsistent with that of {\namesh}.
Binaries with later-type primaries provide a
better match at optical wavelengths, but result in stronger
{\meth} absorption at 1.6 and 2.2~$\micron$ (Figure~\ref{fig_double}).
Similar results were found for alternate L and T dwarf
absolute magnitude/spectral type relations \citep{liu06,meltbinary}.

We therefore find no reasonable combination of normal L and T dwarf 
spectra that can simultaneously 
reproduce the optical and near-infrared spectrum of  
{\namesh}, confirming the conclusion of \citet{fol07} that this source
is likely to be single.  It is of course possible that {\namesh}
is a binary with peculiar components.  
However, this scenario is less compelling
than that in which {\namesh} is a solitary peculiar L dwarf.

\subsubsection{Subsolar Metallicity?}

A common explanation for the spectral peculiarities of blue L dwarfs is
that their atmospheres are metal-depleted,
causing a relative enhancement in collision-induced {\hh} absorption
that preferentially suppresses flux at $K$-band \citep{lin69,sau94,bor97}.
This, along with a general reduction in metal opacity at shorter wavelengths,
results in bluer $J-K$ colors.  Indeed,
blue near-infrared colors are common
for metal-poor M- and L-type subdwarfs \citep{bes82,leg00,me0532}.
Low temperature metal-poor dwarfs also tend to exhibit stronger metal hydride bands
and single metal lines due to the greater relative 
reduction in metal oxide absorption (e.g., \citealt{mou76}).  This trend is
also consistent with enhanced FeH and $J$-band alkali line
absorption observed in the near-infrared spectra
of {\namesh} \citep{fol07} and other blue L dwarfs.  
In addition, L subdwarfs exhibit surprisingly enhanced TiO absorption,
unexpected for a metal-depeleted atmosphere but consistent with 
reduced condensate formation \citep{me0532,giz06,rei06,megmos}.  
{\namesh} appears to exhibit this trait as well (Figure~\ref{fig_optspec}).

However, it is clear that the sources shown in Figure~\ref{fig_blue}
are not as metal-poor as currently known
L subdwarfs, given that the latter 
have far bluer
near-infrared colors ($J-K_s \approx 0$)
and more peculiar optical and near-infrared
spectral morphologies \citep{giz06,megmos}.
Furthermore, the {\nai}, {\rbi} and {\csi} lines in the 7800--8600~{\AA} spectral band are similar in strength to those of 
both 2MASS~J2224-0158 and 2MASS~J1507-1627
(Figure~\ref{fig_alkalines}), making it unlikely that
{\namesh} is significantly metal-poor relative to these sources.

Can the spectral peculiarities seen in blue L dwarf spectra
nevertheless be the result of modest subsolar metallicities; 
e.g., [M/H] $\sim$ -0.5?  To address this question,
we examined trends in the most recent theoretical spectral models
from \citet{bur06}.
While these models do not as yet reproduce the near-infrared spectra
of L dwarfs in detail (cf., \citealt{bur06,cus07}), trends 
as a function of metallicity can be examined and compared to
the deviations observed between normal 
and blue L dwarfs.
Figure~\ref{fig_modelzcomp} illustrates this, comparing two normalized and smoothed
condensate cloud models, both assuming {\teff} = 1700~K, 
{\logg} = 5.5 (cgs) and a modal grain size $a_0$ = 100~$\micron$, 
but differing in metallicity: [M/H] = 0 and -0.5.  
Consistent with the arguments above,
the lower metallicity model exhibits both a 
bluer spectral energy distribution and enhanced 0.99~$\micron$
FeH absorption, as observed in the blue L dwarfs.  However, no enhancement in  
the 1.4~$\micron$ {\wat} band is seen, as its opacity is uniformly
reduced across most of the near-infrared spectral region.  
This inconsistency suggests that subsolar
metallicity alone cannot explain the spectral peculiarities of
blue L dwarfs, even if this is a characteristic trait of such sources.

\subsubsection{High Surface Gravity?}

Surface gravity influences the emergent spectral energy distribution
of a late-type dwarf by modulating the photospheric gas pressure,
affecting both pressure-sensitive features and gas/condensate chemistry.
Spectral signatures of low surface gravity, including weakened
alkali lines, enhanced metal oxide absorption and reduced {\hh}
absorption (resulting in redder near-infrared colors),
have all been used to identify and characterize
young brown dwarfs  
(e.g., \citealt{luh99,mcg04,kir06,all07}).
As blue L dwarfs tend to have opposing spectral 
peculiarities, it is reasonable to consider that these
sources may have high surface gravities, a result of being both
older and more massive than their equivalently classified
counterparts.

As discussed in $\S$~3.4, there is kinematic evidence to support this idea.
{\namesh}, 2MASS~J1300+1921 and 2MASS~J1721+3344 
all have estimated {\vtan} $\gtrsim$ 100~{\kms}
\citep{giz00,cru07}, at the 3$\sigma$ tail of the 
L dwarf {\vtan} distribution of \citet{sch07}.
\citet{kna04} and \citet{cru07} have argued that the large
space velocities of blue L dwarfs indicate that they may be 
members of the old disk or thick disk populations and, as such, are older
and more massive than the average field L dwarf.
The 2~Gyr lower age limit of {\namesh} based on the absence of {\lii} 
absorption in its optical spectrum is further evidence that this source 
is relatively old and has a high surface gravity.

However, surface gravity effects alone also fail to explain the 
spectral peculiarities of {\namesh} and other blue L dwarfs.
Figure~\ref{fig_modelgcomp} illustrates trends in surface gravity
between {\logg} = 5.0 and 5.5 for the condensate cloud models
of \citet{bur06} and M.\ Marley et al.\ (in preparation), assuming {\teff} = 1700~K,
solar metallicity, and baseline cloud parameters ($a_0$ = 100~$\micron$
and {\fsed} = 3; see \citealt{cus07}).
An increase
in surface gravity in the Burrows et al.\ models results in similar
qualitative trends as decreased metallicity, namely
bluer near-infrared colors and somewhat stronger FeH and alkali line 
absorption.  However, increasing the surface gravity 
from {\logg} = 5.0 to 5.5 does not appear to change the depth of 
the 1.4~$\micron$ {\wat} band in any way.  
The Marley et al.\ models do show a change in 
{\wat} band strength with higher surface gravity, but in the opposite
sense as observed in the blue L dwarfs; the absorption becomes weaker.
The reduction in the {\wat} band contrast appears to be due to increased
condensate opacity in the higher surface gravity models, 
affecting the flux peaks but not the deep molecular bands
\citep{ack01,mar02}.
Stronger condensate absorption at $J$-band counteracts the increased
{\hh} absorption at $K$-band, such that $J-K$ colors are only
modestly affected by changes in surface gravity in these models
(see Figure~8 in \citealt{kna04}).
Hence, while blue L dwarfs like {\namesh} may have higher surface gravities than
normal field dwarfs, this trait alone does not explain the 
spectral peculiarities observed.

\subsubsection{Thin Condensate Clouds?}

A fourth possibility is that the condensate clouds of blue L dwarfs
are somehow thinner or less opaque than those of normal field L dwarfs
\citep{kna04,chi06,cru07,leg07}.  Reduced condensate
opacity in the 1~$\micron$ spectral region allows other
features such as FeH and alkali line absorption to appear
stronger at these wavelengths, much as reduced metal oxide absorption 
allows metal hydride bands and alkali lines to emerge in
the red optical spectra of L dwarfs \citep{kir00}
and late-type M subdwarfs \citep{megmos}.
Reduced condensate
opacity also increases the contrast between the $J$-band peak and the
base of the 1.4~$\micron$ {\wat} band, producing a deeper 
feature; and between the $J$ and $K$-band peaks (the latter
dominated by {\hh} opacity),
resulting in bluer near-infrared colors.  Many
of the spectral peculiarities observed in blue L dwarfs can
be qualitatively explained by a reduction in condensate opacity.

Theoretical spectral models quantitatively confirm these trends as well.
Figure~\ref{fig_modelfcomp} compares models from \citet{bur06}
and M.\ Marley et al.\ (in preparation) for {\teff} = 1700~K, {\logg} = 5.5 and solar
metallicity, but with different treatments for the properties of the
condensate cloud layers.  For the Burrows et al.\ models, we
compare different
values for the condensate modal grain size, $a_0$ = 30~$\micron$ versus
100~$\micron$.  Larger grain sizes for a given total condensate mass
corresponds to fewer grains and smaller total opacity 
(see Figure~6 in \citealt{bur06}), resulting in a
bluer near-infrared spectral energy distribution and 
stronger absorption features at the $J$-band flux peak.
In particular, the 1.4~$\micron$
{\wat} band is clearly enhanced in the larger-grain model.
For the Marley et al.\ models, we compared different values for the
{\fsed} parameter, which describes the 
efficiency of condensate sedimentation.
Larger values of {\fsed} correspond to both thinner clouds and
larger mean particle sizes \citep{ack01}.
The trends are qualitatively similar to the Burrows et al.\ models:
stronger atomic and molecular gas features and a 
bluer near-infrared spectral energy distribution as seen
in the blue L dwarfs.  

\citet{kna04} have previously demonstrated that the {\em colors}
of blue L dwarfs can be reproduced with models with thinner clouds 
(higher values of {\fsed}).
Figure~\ref{fig_modelfcomp} demonstrates that the {\em spectra} of these
sources can be reproduced with thin cloud
models as well.  
The {\fsed} = 4, {\teff} = 1700~K and {\logg} = 5.5 model
of Marley et al.\ provides an excellent match to the overall 
spectral energy distribution 
of {\namesh}, including its blue color and deep FeH and {\wat}
absorption bands.  
In contrast, most midtype L dwarf spectra are adequately 
reproduced assuming {\fsed} = 2--3 \citep[M.\ Marley et al.\ in preparation]{kna04,cus07}.
\citet{cus07} have also found that models with 
progressively thinner clouds fit 
progressively bluer objects across the L dwarf/T dwarf transition.

Further evidence that cloud properties play a significant
role in the colors of L dwarfs can also be deduced from 
mid-infrared spectroscopy.  \citet{cus06} have
recently reported a tentative identification of the Si-O stretching mode
in the mid-infrared spectra of three midtype L dwarfs.  This feature
arises from small silicate grains in the photospheres 
of cloudy L dwarfs.  If differences in the 
colors of L dwarfs are caused by condensate cloud effects, 
they should be correlated with the 
strength of the Si-O feature.  This appears to be the case.
Figure~\ref{fig_irs} compares the mid-infrared
spectra of two sources from the \citet{cus06} study,
the L4.5 2MASS J2224-0158 and the L5 2MASS~J1507-1627, whose optical
and near-infrared spectra are
also shown in Figures~\ref{fig_optspec} and~\ref{fig_nirspec}.
These sources have $J-K_s$ colors that differ by over 0.5~mag,
and are $\sim$0.2--0.3~mag redder and bluer than the average midtype
L dwarf, respectively \citep{kir00}.  The 10~$\micron$ feature noted by \citet{cus06}
is clearly weaker in the bluer L dwarf, 
consistent with the interpretation of thinner and/or larger
grained clouds.  By analogy, we expect this feature
in blue L dwarfs such as {\namesh} to be weaker still.
This prediction can be tested with future observations.

The presence of thin (or large-grained) uniform condensate clouds therefore
provides an adequate explanation for the spectral peculiarities
of {\namesh} and other blue L dwarfs.  
\citet{fol07} have also proposed a somewhat different
interpretation of cloud properties in {\namesh}:
that the apparently 
reduced condensate opacity arises from holes in an otherwise thick cloud layer.
This draws from an idea set forth by \citet{ack01} and \citet{mecloud} to
explain the rapid disappearance of cloud opacity across the
L dwarf/T dwarf transition. Indeed, \citet{fol07} suggest that
{\namesh} is itself an L/T transition object (consistent with its late
near-infrared spectral type) that may be ``crossing over'' at
an early stage, perhaps due to reduced metallicity (see also \citealt{bur06}).  The midtype L dwarf optical spectral morphology
of this source, much earlier than the optical spectra of any T dwarf
observed to date \citep[J.\ D.\ Kirkpatrick et al., in preparation]{me03opt},
and the absence of {\meth} absorption at 2.2~$\micron$, a common feature of
L9 dwarfs \citep{geb02}, argues against this hypothesis.  On the other hand,
the blue L dwarf 2MASS~J1300+1912 exhibits strong photometric 
variability \citep{gel02,mai05} likely due to cloud structure, including
perhaps cloud holes.  Monitoring of {\namesh} may provide insight into
the ``cloud hole'' interpretation of this source and blue L dwarfs in general.
In any case, the basic premise of \citet{fol07}, that {\namesh} has reduced
condensate opacity, is in agreement with our analysis.

\subsection{Improved Near-Infrared Classification of Blue L Dwarfs}

\begin{figure}
\epsscale{1.0}
\plotone{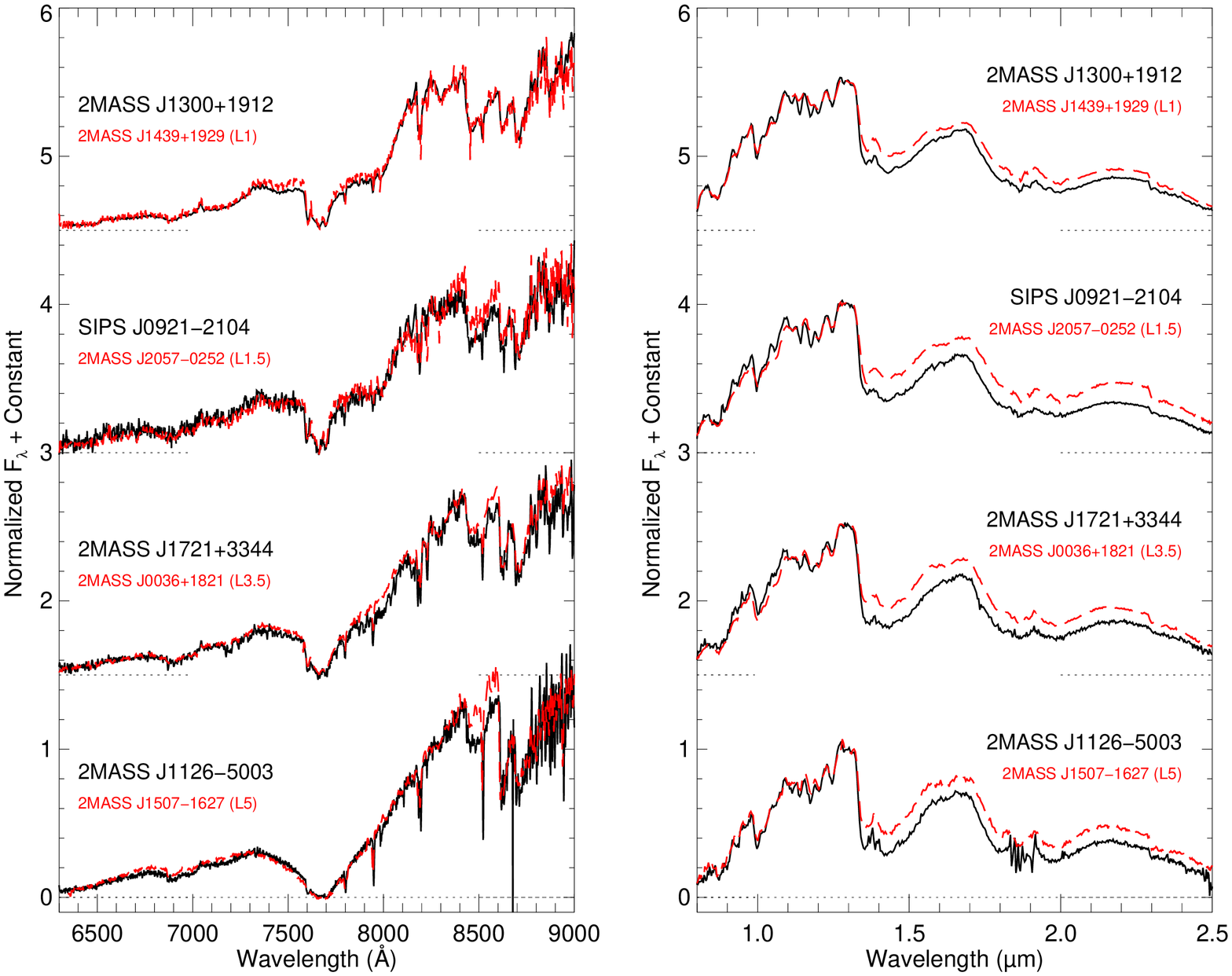}
\caption{Comparison of optical (left panel, 6300--9000~{\AA}) 
and near-infrared (right panel, 0.9--2.4~$\micron$)
spectra of the four blue L dwarfs
2MASS~J1300+1912, SIPS~J0921-2104,
2MASS~J1721+3344 and
{\namesh} (black lines, from top to bottom).  These are compared to field L dwarf
spectral templates
2MASS~14392836+1929149 (L1; \citealt{kir99}), 
2MASS~J20575409-0252302 (L1.5; \citealt{cru03}),
2MASS~J00361617+1821104 (L3.5; \citealt{rei00})
and 2MASS~J1507-1627 (L5; \citealt{rei00}; red dashed lines).  
All spectra are normalized at 8200~{\AA} (left panel)
or 1.28~$\micron$ (right panel)
and offset by constants (dotted lines).
Note that the blue L dwarfs show reasonable agreement with their
spectral comparison sources up to the $\sim$1.3~$\micron$ {\wat}
band, but are depressed (to varying degrees) at longer wavelengths.
\label{fig_blue}}
\end{figure}

\begin{figure}
\epsscale{0.8}
\plotone{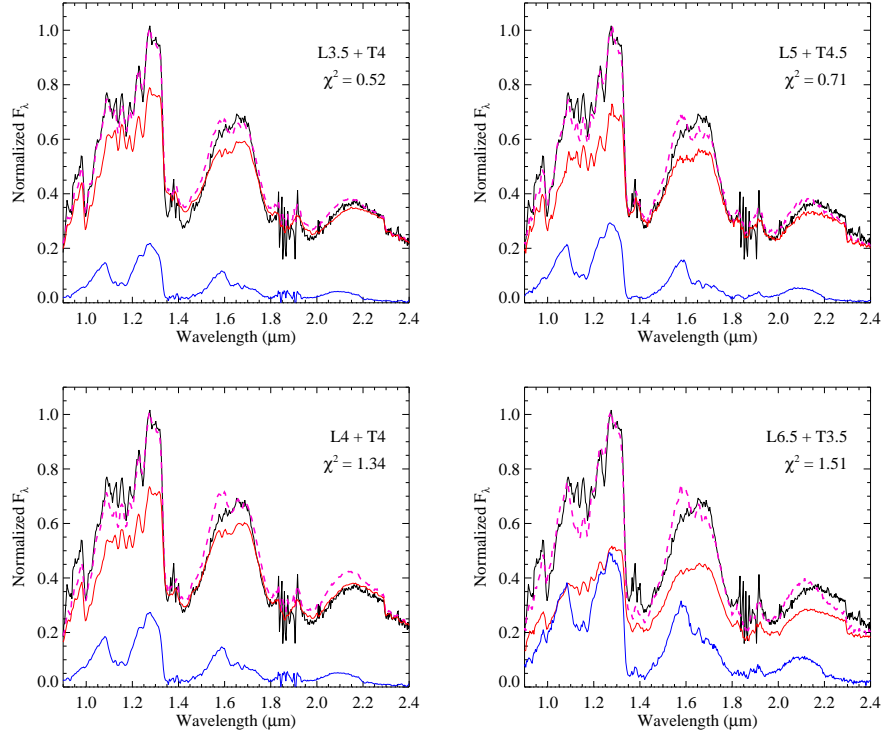}
\caption{Best-fitting binary brown dwarf templates to the near-infrared
spectrum of {\namesh}.  Data for the source (black lines) and spectra of the binary composites (purple dashed lines) are normalized at 1.27~$\micron$.
Primary (red solid lines) and secondary (blue solid lines) template
spectra are scaled to their relative contributions to the composite spectra.
The spectral types of the components (optical for L dwarfs, near-infrared for T dwarfs) are indicated in the upper right corner of each panel, along with
$\chi^2$ deviations.
\label{fig_double}}
\end{figure}

\begin{figure}
\epsscale{0.8}
\plotone{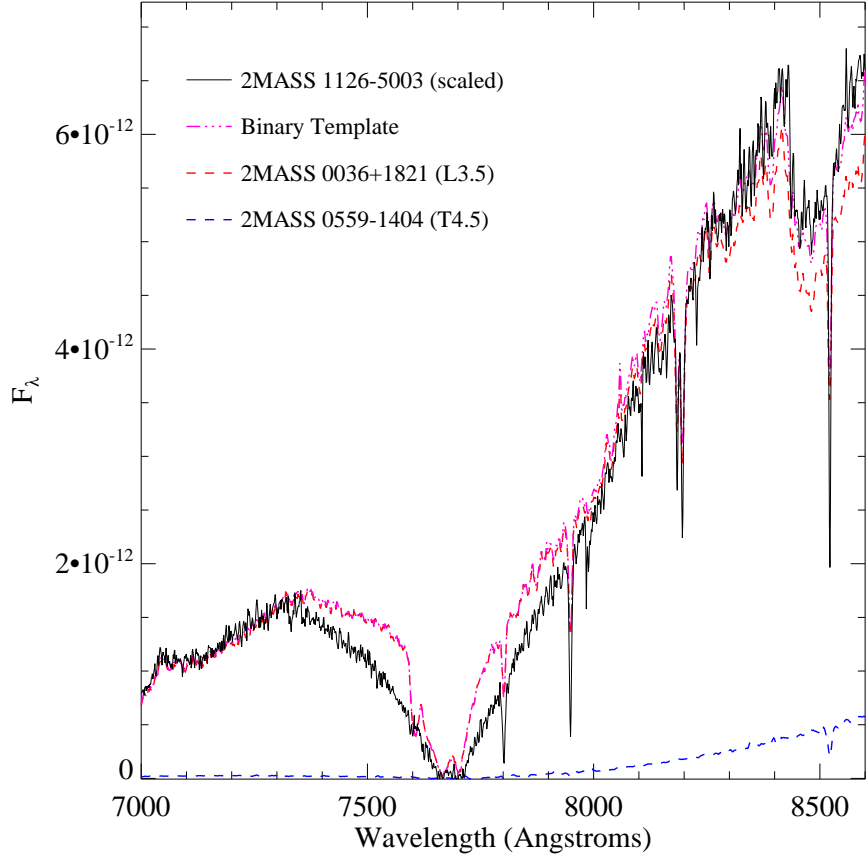}
\caption{Comparison of a composite binary spectrum (purple dashed line)
composed of L3.5 2MASS~J0036+1821 (red triple-dot-dashed line) 
and T4.5 2MASS~J0559-1404 (blue triple-dot-dashed line) template
spectra, to optical data for {\namesh} (black solid line).  The templates are scaled to their $M_{I_c}$ magnitudes as measured by \citet{dah02}, and the spectrum of
{\namesh} is scaled to provide a best fit to the composite spectrum.  
\label{fig_doubleopt}}
\end{figure}

\begin{figure}
\epsscale{0.8}
\plotone{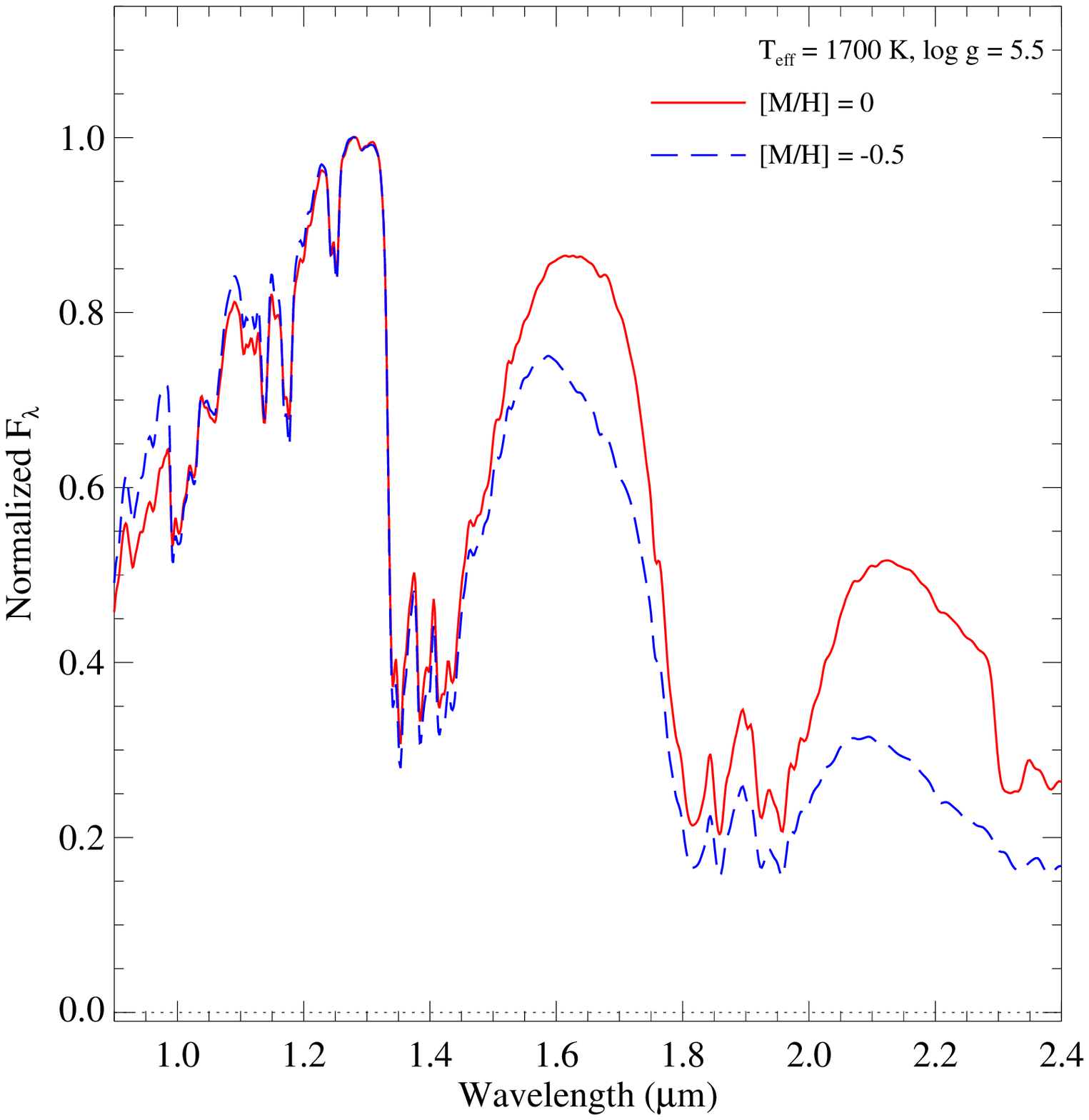}
\caption{Comparison of metallicity effects in the spectral models 
of \citet{bur06}.  Both models shown assume
{\teff} = 1700~K, {\logg} = 5.5 (cgs) and a modal condensate
particle size $a_0$ = 100~$\micron$, but differ in metallicity 
(red solid line: [M/H] = 0, blue dashed line: [M/H] = -0.5).  
The model spectra have been
smoothed to a resolution of {\ldl} = 120
using a Gaussian kernel, similar to that of the SpeX data,
and are normalized at 1.27~$\micron$.
\label{fig_modelzcomp}}
\end{figure}

\begin{figure}
\epsscale{1.0}
\plottwo{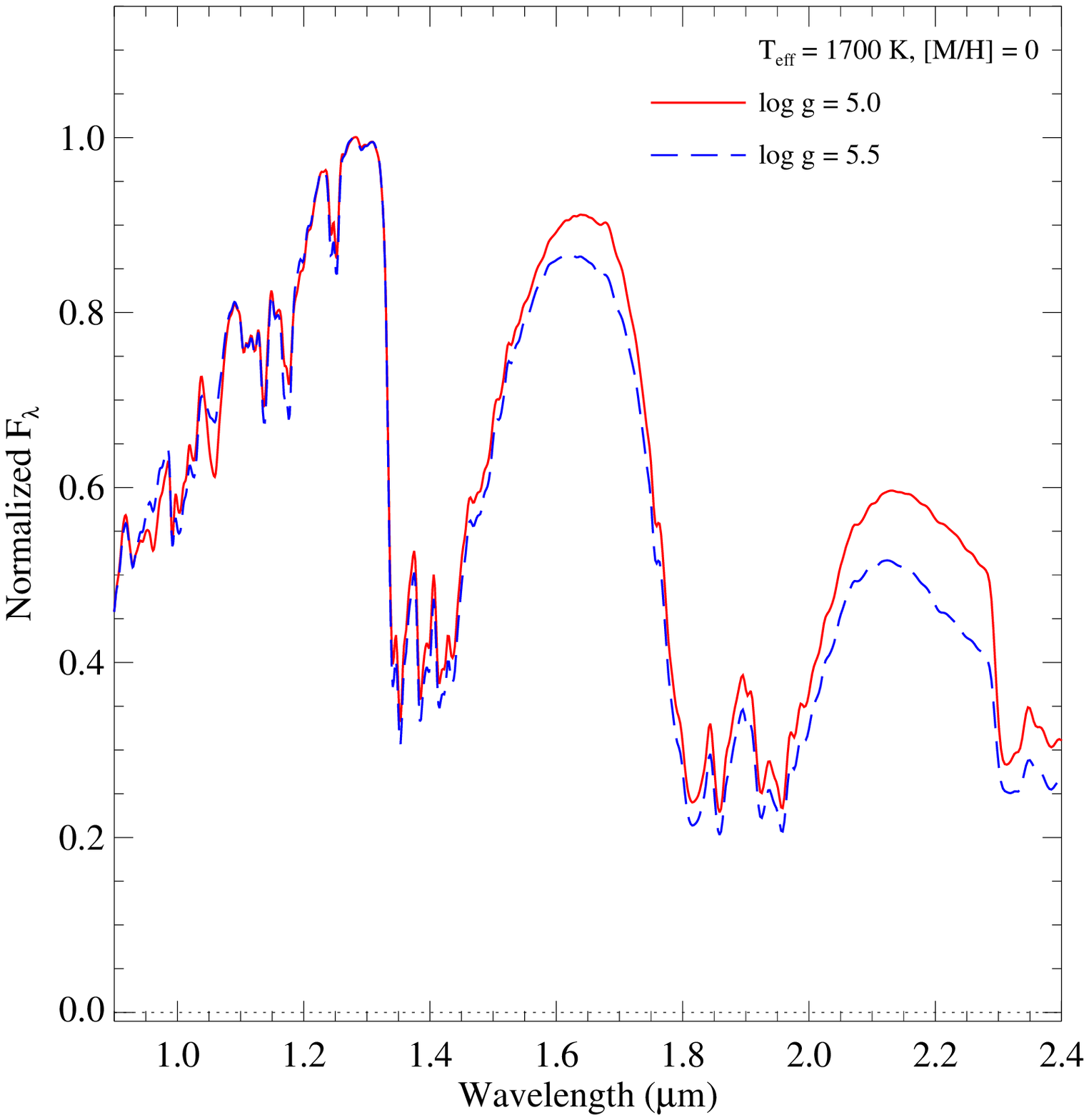}{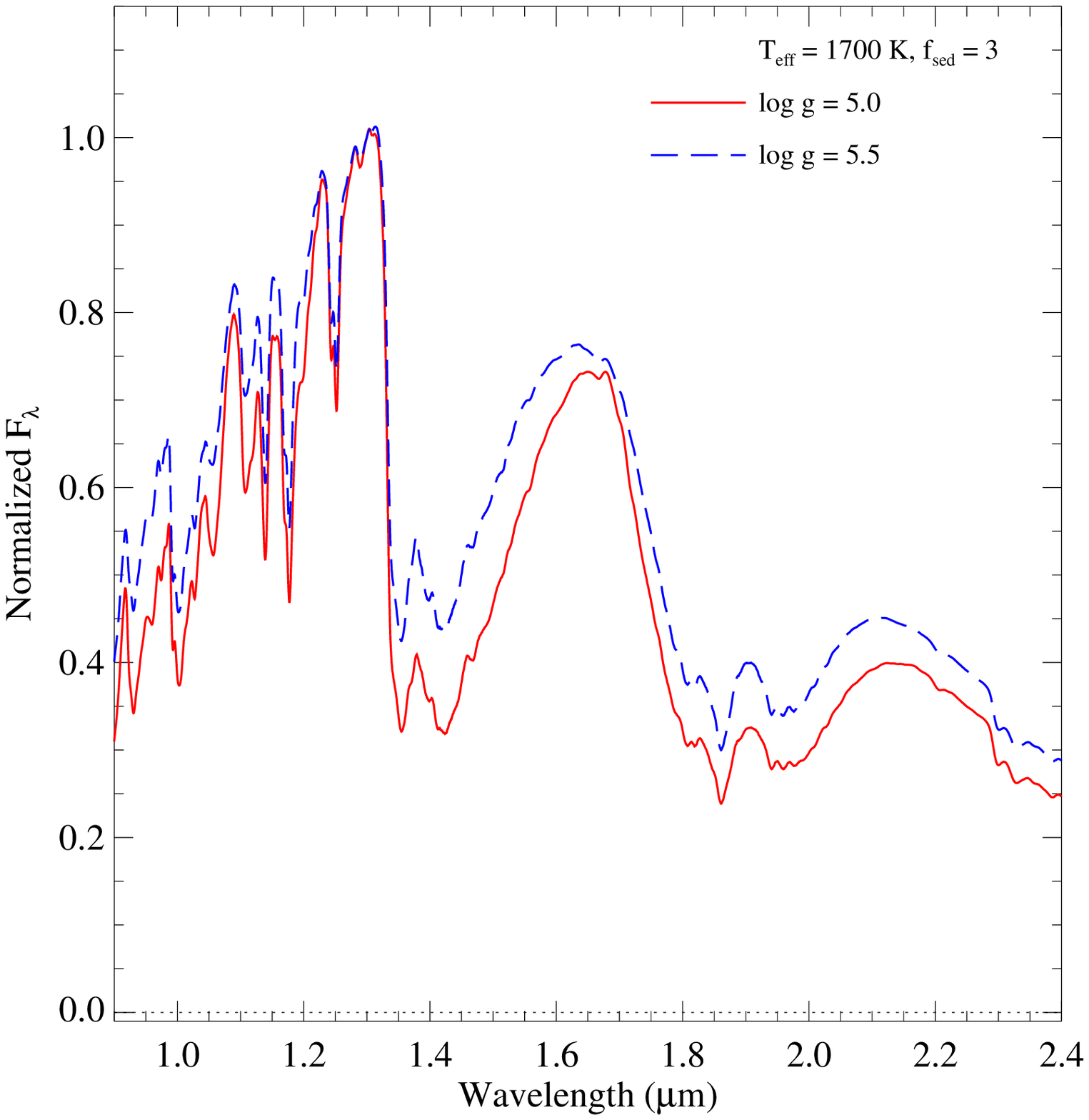}
\caption{Comparison of gravity effects in the 
spectral models of \citet[left]{bur06}
and M.\ Marley et al.\ (in preparation; right).  All models assume 
{\teff} = 1700~K and solar metallicity but differ in surface gravity 
(red solid lines: {\logg} = 5.0, blue dashed lines: {\logg} = 5.5).
All model spectra have been
smoothed to a resolution of {\ldl} = 120
using a Gaussian kernel 
and are normalized at 1.27~$\micron$.
\label{fig_modelgcomp}}
\end{figure}

\begin{figure}
\epsscale{1.0}
\plottwo{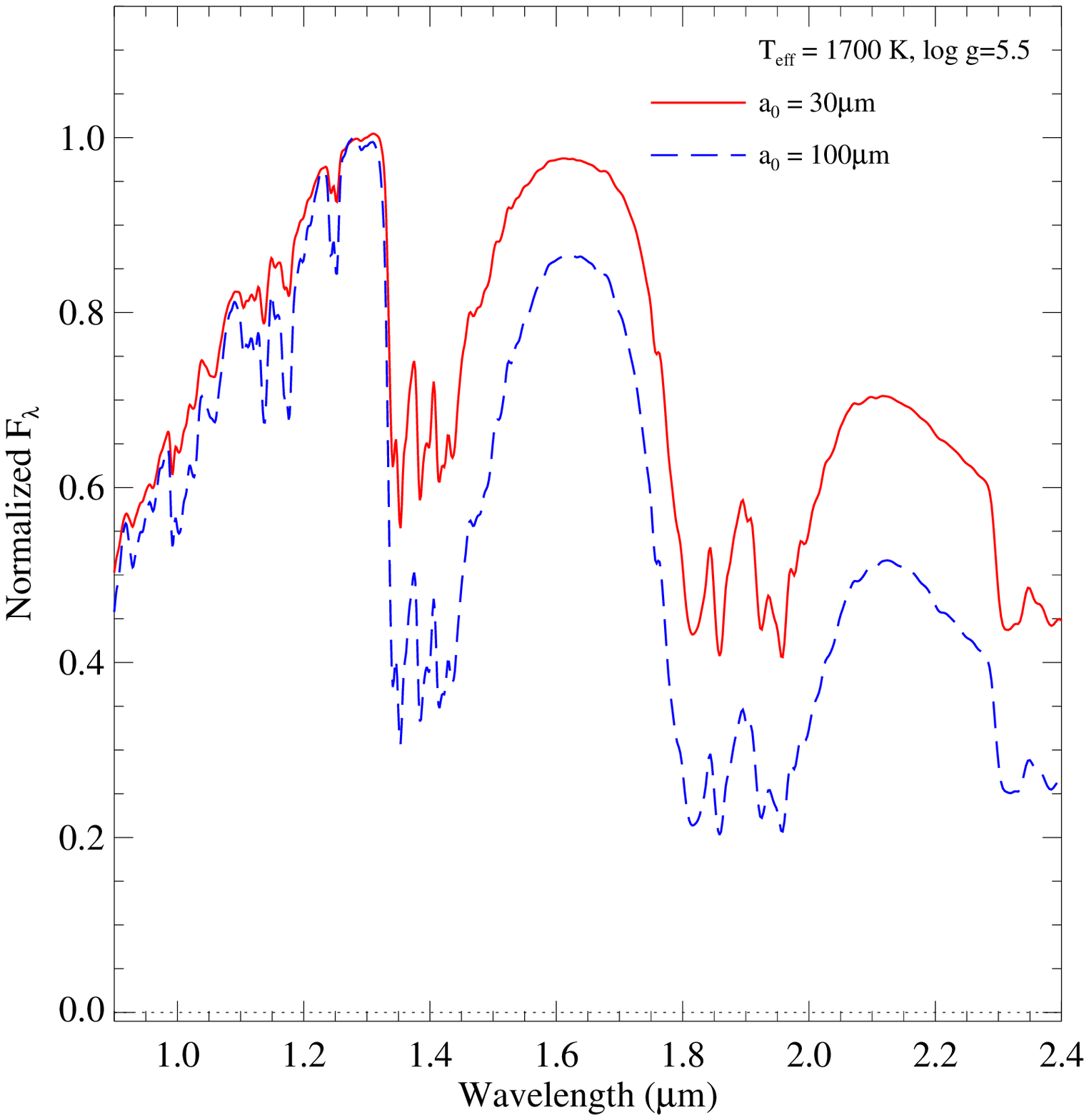}{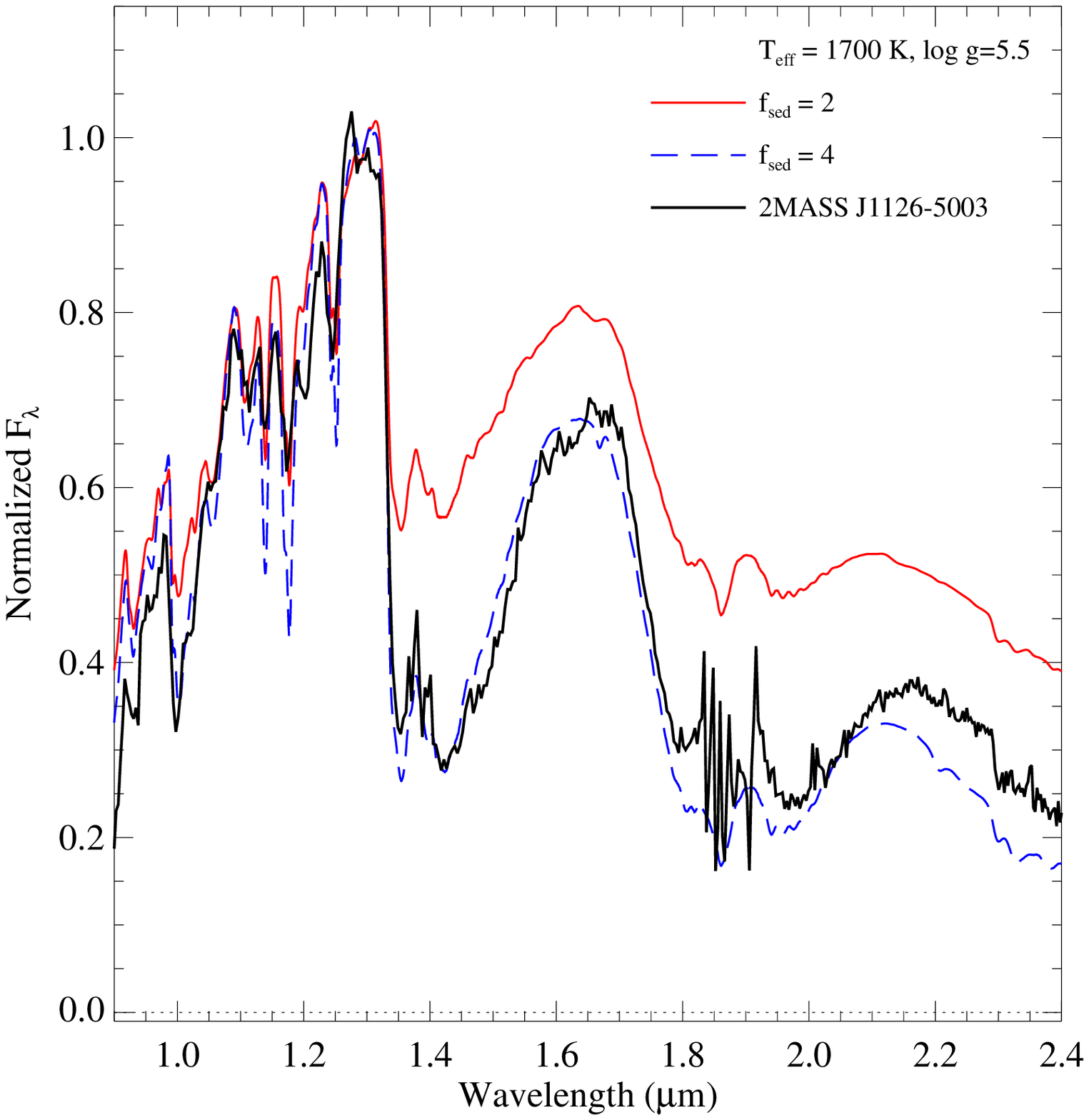}
\caption{Comparison of condensate cloud effects in the 
spectral models of \citet[left]{bur06}
and M.\ Marley et al.\ (in preparation; right).  All models assume 
{\teff} = 1700~K, solar metallicity, and {\logg} = 5.5.
The Burrows et al.\ models differ in their modal condensate particle
size, with $a_0$ = 30~$\micron$ (red solid line) and 100~$\micron$ (blue 
dashed line)
shown.  The Marley et al.\ models differ in the assumed sedimentation
efficiency, with {\fsed} = 2 (red solid line) and 4 (blue dashed line) shown.
Also shown in the right panel is the spectrum of {\namesh}
(black solid line), which
shows adequate agreement with the {\fsed} = 4 model of Marley et al.
Model spectra have been
smoothed to a resolution of {\ldl} = 120
using a Gaussian kernel 
and are normalized at 1.27~$\micron$.
\label{fig_modelfcomp}}
\end{figure}

\begin{figure}
\epsscale{1.0}
\plotone{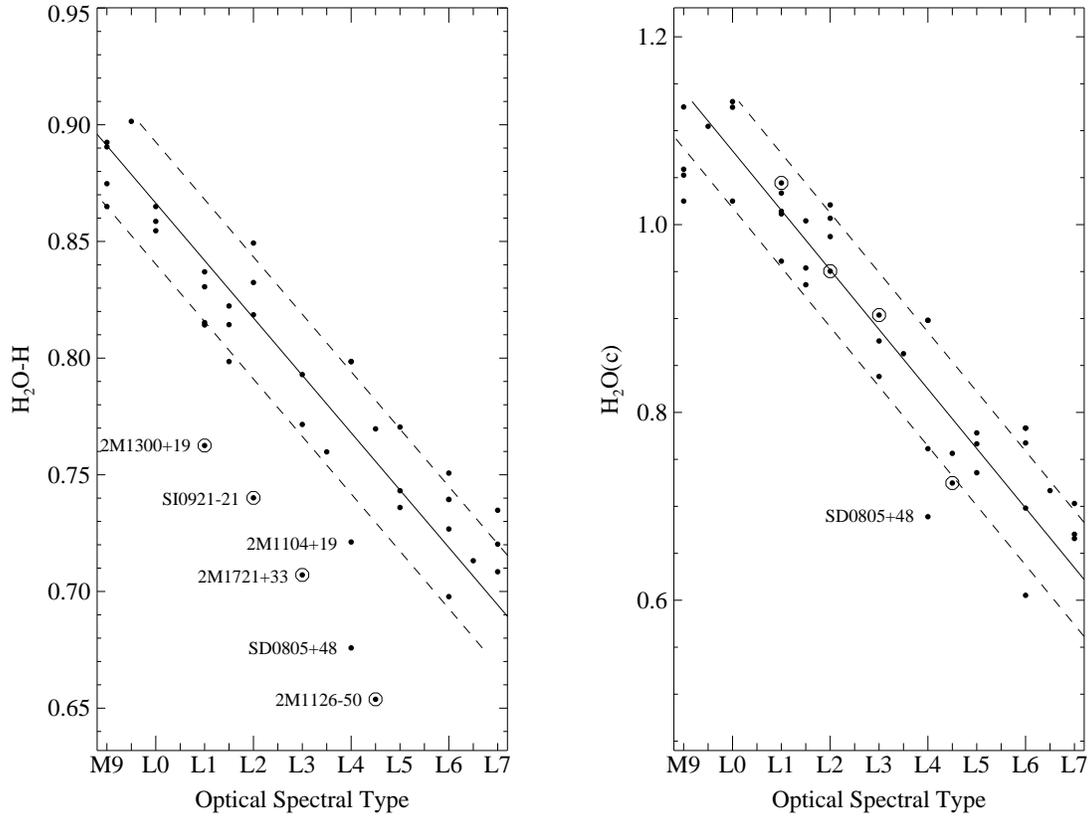}
\caption{Values for the
1.4~$\micron$ {\wat} spectral ratios {\wat}-H (left) and {\wat}(c) (right)
as a function of optical spectral type 
for 39 unresolved, non-peculiar and optically classified 
field late-M and L dwarfs (dots) and
the four blue L dwarfs in Figure~\ref{fig_blue} (circled dots).
Linear fits for the normal field dwarfs are indicated by solid lines,
with dashed lines indicating 1$\sigma$ scatter (roughly 1~subtype).  
The blue L dwarfs clearly stand out in the left panel due to their 
enhanced 1.4~$\micron$ {\wat} absorption,  
while the color-corrected {\wat}(c) ratio yields subtypes consistent
with their optical types within the scatter.
The additional L4 outliers in the left panel 
are SDSS J080531.84+481233.0 \citep{haw02,kna04}
and 2MASS J11040127+1959217 \citep{cru03}, which have also have blue $J-K_s$
colors.  The former may be an unresolved binary \citep{me0805}.
\label{fig_nirind}}
\end{figure}

\begin{figure}
\epsscale{0.9}
\plotone{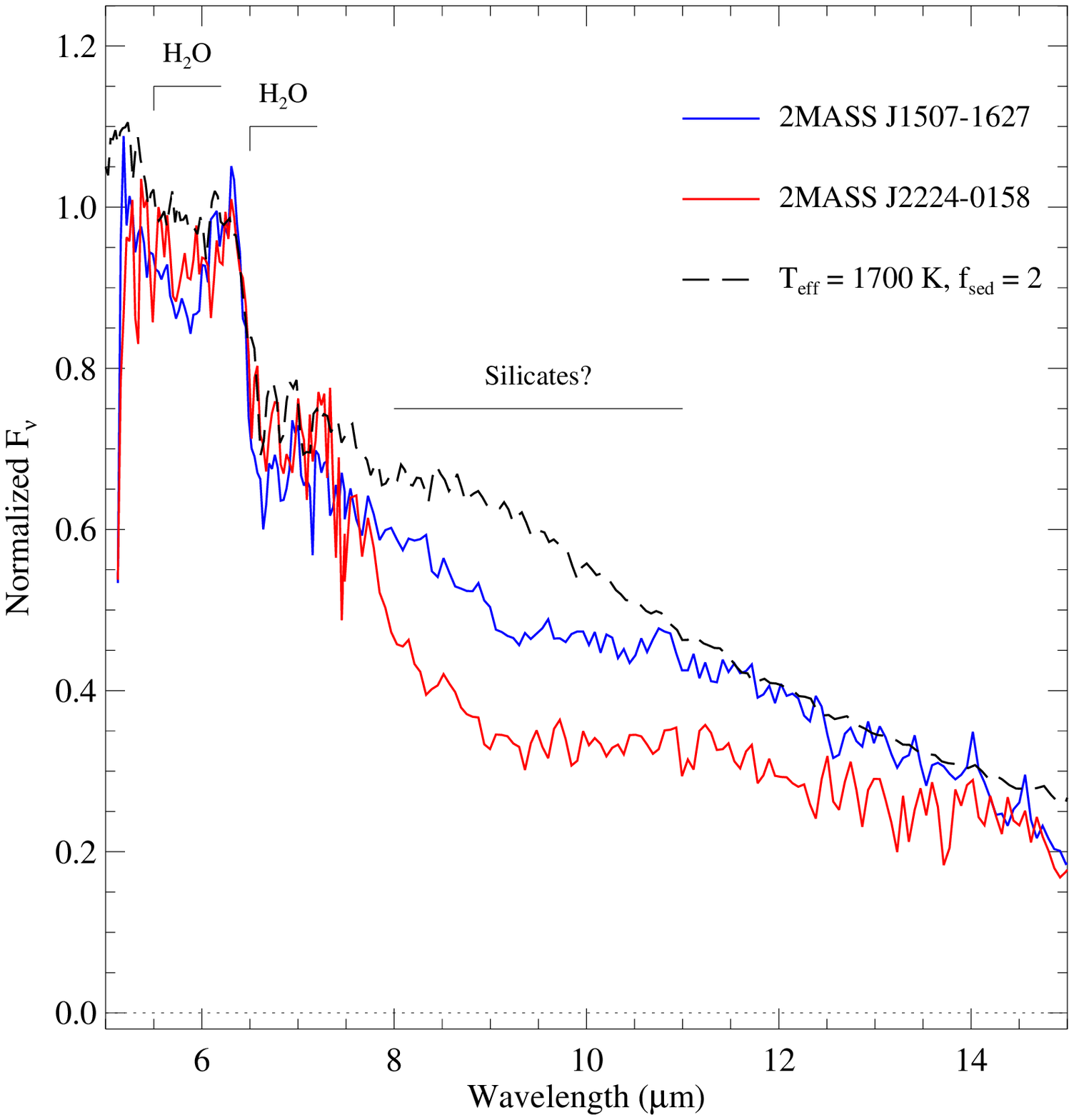}
\caption{Mid-infrared spectra obtained with the {\em Spitzer}
Infrared Spectrograph \citep{hou04} of the L dwarfs
2MASS~J2224-0158 (L4.5; red line) and
2MASS~J1507-1627 (L5; blue line)
from \citet{cus06}.  These sources are
$\sim$0.2--0.3~mag bluer and redder in $J-K_s$
than the average midtype L dwarf,  respectively
\citep{kir00}.  
A broad absorption feature apparently present
in the 8--11~$\micron$ region of both spectra 
has been tentatively identified
as arising from the Si-O stretch mode associated with
condensate species.  This feature is clearly
weaker in the bluer and possibly less cloudy L dwarf 2MASS~J1507-1627.
Also shown is a {\teff} = 1700~K, 
{\logg} = 5.0, {\fsed} = 2 spectral model
from M.\ Marley et al.\ (in preparation.) that does not exhibit this feature
(see discussion in \citealt{cus06}).
All spectra are normalized
at 6.4~$\micron$.
\label{fig_irs}}
\end{figure}

We now readdress the issue of the
discrepancies between optical and near-infrared classifications 
of blue L dwarfs like {\namesh}.  These arise largely from
the enhanced 1.4~$\micron$ {\wat} band, which as illustrated in Figure~\ref{fig_modelfcomp} 
is highly sensitive to condensate cloud properties (see also \citealt{ste03}).
Since clouds also influence the near-infrared colors of L dwarfs, one possible
way of reconciling these optical and near-infrared types is to use a
color-corrected {\wat} index, analogous to the color-independent
indices used to classify young reddened M and L dwarfs (e.g., \citealt{wil99,all07}).
We constructed a ``hybrid'' index:
\begin{eqnarray}
H_2O(c)  & \approx & [H_2O-A]/[H/J] \\
 & \equiv & \left[ \frac{\int{F_{1.33-1.35}}}{\int{F_{1.58-1.60}}}\right] , \\ \nonumber
\end{eqnarray}
where {\wat}$-A$ is defined in \citet{rei01} and $H/J$
is defined in \citet{meclass}.  Figure~\ref{fig_nirind}
compares this ratio with another ratio sampling the 1.4~$\micron$
band ({\wat}$-H$ from \citealt{meclass2}) as a function of optical
spectral type (SpT) for a sample of SpeX prism data of unresolved and non-peculiar 
late-type M and L dwarfs 
(data from \citealt{meltbinary} and K.\ Cruz et al.\ in preparation)
and the five blue L dwarfs
shown in Figure~\ref{fig_blue}.
Linear fits to the normal M and L dwarfs yield the relations
\begin{equation}
SpT = 47.70 - 43.67[H_2O-H] \label{eqn_ind1}
\end{equation}
\begin{equation}
SpT = 27.41 - 16.17[H_2O(c)] \label{eqn_ind2}
\end{equation}
with a scatter of 0.9 subtypes for both.
The blue L dwarfs clearly stand apart 
in the spectral type/{\wat}$-H$ comparison, 
and application of Equation~\ref{eqn_ind1} yields 
near-infrared
spectral types that are 3.5-4 subtypes later
than their optical types (Table~\ref{tab_classblue}).  However, the subtypes
inferred using Equation~\ref{eqn_ind2} for the four blue L dwarfs
shown in Figure~\ref{fig_blue}
are consistent with their optical types to
within one subytpe.  Note that SDSS~J0805+4812 still stands
apart from the locus of spectral type versus {\wat}(c), probably because
its peculiarities arise from unresolved
multiplicity as opposed to cloud effects.
We advocate use of the {\wat}(c) ratio, along with 
ratios sampling the longer wavelength features
(e.g., K1 or {\meth}-2.2~$\micron$), as cloud-independent estimators
for the optical spectral types of single L dwarfs.  

As pointed out in $\S$~3.3, there is as yet no formal near-infrared classification
scheme for L dwarfs.  Existing practice --- tying near-infrared indices
to optical spectral types --- ignores the fact that secondary  physical parameters
such as the character of condensates and condensate clouds can modify the optical
and near-infrared spectra of L dwarfs in different ways.  
Indeed, blue L dwarfs stand out as peculiar largely because their near-infrared
spectra do not conform to the morphologies expected for their optical types.
Future efforts at extending the existing L dwarf optical classification scheme
to encompass near-infrared spectral morphologies will likely require 
consideration of additional classification parameters that take into account
secondary effects, much as luminosity classes delineate surface gravity effects in stars.
(for further discussion of these issues, see \citealt{kir05}).
The definition of a multi-dimensional near-infrared classification scheme
for L dwarfs is clearly beyond the scope of this study.  We simply point out that
blue L dwarfs such as {\namesh} are likely to serve as
useful standards for delineating future ``cloud classes'' amongst L dwarfs.

\section{Discussion}

Our analysis in $\S$~4.2 leads us to conclude that the spectral 
peculiarities of {\namesh} and other blue L dwarfs have their
immediate cause in condensate
cloud effects, specifically the presence of thin, patchy or large-grained
condensate clouds at the photosphere.  Subsolar metallicities and 
high surface gravities in of themselves
cannot reproduce the observed spectral peculiarities of these sources. 
However, it is clear that these latter physical properties 
must play a role in determining the cloud characteristics of blue L dwarfs.
Lower metallicities reduce the metal species
available to form condensates, resulting in less condensate material
overall.  Higher surface gravities may increase the sedimentation rate
of condensate grains, potentially resulting in thinner clouds.
The large tangential velocities 
and absence of {\lii} absorption in the three blue
L dwarfs {\namesh}, 2MASS~J1300+1921 and 2MASS~J1721+3344 support the
idea that these sources may be relatively old and possibly slightly metal-poor. 
However, the influence of other physical parameters on condensate
cloud properties must also be considered, including rotation rates, 
vertical upwelling rates (e.g., \citealt{sau06}) and possibly 
magnetic field strengths.  

An assessment of how these fundamental physical
parameters influence the properties of condensate
clouds in low-temperature atmospheres is the subject of ongoing theoretical investigations
(e.g., \citealt{hel01,woi03}; M.\ Marley, in preparation).
Empirical studies are also necessary, particularly those focused on 
well-characterized samples of blue 
(and red) L dwarfs.
To this end, Table~\ref{tab_blue} lists all
blue L dwarfs currently reported in the literature.  
We anticipate that this
list will grow as near-infrared spectroscopic follow-up of L dwarfs continues.

\acknowledgements
The authors thank Adam Burrows, Mark Marley and Didier Saumon for
providing spectral models for our analysis and comments on the original
manuscript, and Michael Cushing for making
available his Spitzer IRS data.  We would also like to thank 
telescope operator Bill Golisch
and instrument specialist John Rayner at IRTF, and telescope operator
Hern\'{a}n Nu\~{n}ez at Magellan for their assistance during the
observations.   Additional appreciation goes to our anonymous referee for 
her/his prompt review. This publication makes
use of data from the Two Micron All Sky Survey, which is a joint
project of the University of Massachusetts and the Infrared
Processing and Analysis Center, and funded by the National
Aeronautics and Space Administration and the National Science
Foundation. 2MASS data were obtained from the NASA/IPAC Infrared
Science Archive, which is operated by the Jet Propulsion
Laboratory, California Institute of Technology, under contract
with the National Aeronautics and Space Administration.
This research has benefitted from the M, L, and T dwarf compendium housed at DwarfArchives.org and maintained by Chris Gelino, Davy Kirkpatrick, and Adam Burgasser.
K.~L.~C is supported by a NSF Astronomy and
Astrophysics Postdoctoral Fellowship under AST-0401418.
The authors wish to recognize and acknowledge the 
very significant cultural role and reverence that 
the summit of Mauna Kea has always had within the 
indigenous Hawaiian community.  We are most fortunate 
to have the opportunity to conduct observations from this mountain.

Facilities: \facility{IRTF(SpeX); Magellan Clay(LDSS-3)}

\clearpage

\begin{deluxetable}{llll}
\tabletypesize{\footnotesize}
\tablecaption{Astrometry for {\name}. \label{tab_astrometry}}
\tablewidth{0pt}
\tablehead{
\colhead{$\alpha$\tablenotemark{a}} &
\colhead{$\delta$\tablenotemark{a}} &
\colhead{Epoch}  &
\colhead{Catalog} \\
}
\startdata
11$^h$26$^m$42$\fs$65 & -50$\degr$04$\farcm$02$\farcs$4 & 13 Jan 1983 & ESO; SSS \\
11$^h$26$^m$42$\fs$24 & -50$\degr$04$\farcm$01$\farcs$4 & 13 Mar 1985 & UKST; SSS \\
11$^h$26$^m$39$\fs$93 & -50$\degr$03$\farcm$55$\farcs$3 & 06 Apr 1999 & DENIS \\
11$^h$26$^m$39$\fs$91 & -50$\degr$03$\farcm$55$\farcs$0 & 10 May 1999 & 2MASS \\
11$^h$26$^m$39$\fs$89 & -50$\degr$03$\farcm$55$\farcs$3 & 30 May 1999 & DENIS \\
\enddata
\tablenotetext{a}{Equinox J2000 coordinates.}
\end{deluxetable}

\begin{deluxetable}{ll}
\tabletypesize{\footnotesize}
\tablecaption{Equivalent Widths of Optical Lines. \label{tab_ews}}
\tablewidth{0pt}
\tablehead{
\colhead{Line} &
\colhead{EW ({\AA})}  \\
}
\startdata
H$\alpha$ (6563~{\AA}) & $>-$0.5\tablenotemark{a} \\
\ion{Li}{1} (6708~{\AA}) & $<$~0.4\tablenotemark{a} \\
\ion{Rb}{1} (7800~{\AA}) & 5.6$\pm$0.3 \\
\ion{Rb}{1} (7948~{\AA}) & 5.6$\pm$0.3 \\
\ion{Na}{1} (8183/8195~{\AA}) & 5.6$\pm$0.2 \\
\ion{Cs}{1} (8521~{\AA}) & 3.5$\pm$0.2 \\
\enddata
\tablenotetext{a}{1$\sigma$ upper/lower limits.}
\end{deluxetable}

\begin{deluxetable}{llll|llll}
\tabletypesize{\footnotesize}
\tablecaption{Spectral Indices and Classification. \label{tab_indices}}
\tablewidth{0pt}
\tablehead{
\multicolumn{4}{c|}{Optical} &
\multicolumn{4}{c}{Near-infrared} \\
\cline{1-4} \cline{5-8}
\colhead{Index} &
\colhead{Value} &
\colhead{Subtype\tablenotemark{a}} &
\multicolumn{1}{c|}{Ref} &
\colhead{Index} &
\colhead{Value} &
\colhead{Subtype\tablenotemark{a}} &
\colhead{Ref} \\
}
\startdata
CrH-a & 1.79 & L3.5 & 1 & {\wat}-A & 0.46 & L8.5 & 4 \\
Rb-b/TiO-b & 1.59 & L4.5 & 1 & {\wat}-B & 0.57 & L6.5 & 4 \\
Cs-a/VO-b & 1.33 & L4.5 & 1 & K1 & 0.34 & L4.5 & 4,5 \\
Color-d & 11.87 & L5 & 1 & {\wat} 1.5$\micron$ & 1.74 & [L8] & 6  \\
\ion{K}{1} fit & \nodata & L5 & 1 & {\meth} 2.2$\micron$ & 1.02 & [L5.5] & 6 \\
PC3 & 6.58 & [L5]\tablenotemark{b} & 2 & {\wat}-J & 0.74 & [L6] & 7,8 \\
VO7434 & 1.62 & [L6] & 3 & {\wat}-H & 0.65 & [L8] & 7,8  \\
Na8190 & 1.05 & [L3] & 3 & {\meth}-K & 0.98 & [L5] & 7,8 \\
TiO8440 & 0.82 & [L3] & 3 &  {\wat} & 1.47  & [L7.5] & 9 \\
 & & & & {\wat}(c) & 0.72 & [L5.5] & 10 \\
\cline{1-8}
Optical Type & \multicolumn{3}{c|}{L4.5$\pm$0.5} & NIR Type & \multicolumn{3}{c}{L6.5$\pm$2 (pec)} \\
\enddata
\tablenotetext{a}{Subtypes in brackets were not used to compute
the final average type.}
\tablenotetext{b}{Consistent with measurements for DENIS~J1228-1547 
\citep{del97,mrt99}, classified L5 on the \citet{kir99} scheme.}
\tablerefs{(1) \citet{kir99}; (2) \citet{mrt99}; (3) \citet{haw02};
(4) \citet{rei01}; (5)  \citet{tok99}; (6) \citet{geb02};
(7) \citet{meclass2}; (8) \citet{meltbinary};  (9) \citet{all07};
(10) This paper.}
\end{deluxetable}

\begin{deluxetable}{lll}
\tabletypesize{\footnotesize}
\tablecaption{Properties of {\name}. \label{tab_properties}}
\tablewidth{0pt}
\tablehead{
\colhead{Parameter} &
\colhead{Value} &
\colhead{Reference} \\
}
\startdata
$\alpha$\tablenotemark{a} & 11$^h$26$^m$39$\fs$91 & 1 \\
$\delta$\tablenotemark{a} & $-$50$\degr$03$\arcmin$55$\farcs$0 & 1 \\
$\mu$ & 1$\farcs$66$\pm$0$\farcs$03 yr$^{-1}$ & 1,2,3 \\
$\theta$ & 285$\fdg$3$\pm$1$\fdg$6 & 1,2,3 \\
Optical SpT & L4.5 & 3 \\
NIR SpT & L6.5$\pm$2 (pec) & 3 \\
Distance\tablenotemark{b,c}  & 15$\pm$2 pc & 3 \\
$V_{tan}$\tablenotemark{b} & 117$\pm$15 {\kms} & 3 \\
$V_{rad}$ & $46{\pm}9$ {\kms} & 3 \\
($U,V,W$)\tablenotemark{b} & ($85,-98,-6$) {\kms} & 3 \\
{\teff}\tablenotemark{b} & $\approx$1700~K & 4,5 \\
Mass\tablenotemark{b,d} & $>$0.065~M$_{\sun}$ & 3 \\
Age\tablenotemark{b,d} & $>$2~Gyr & 3,6 \\
$R_{ESO}$ & 20.36 mag & 2 \\
$I_{N}$ & 17.60 mag & 2 \\
$I$ & 17.80$\pm$0.15 mag & 7 \\
$J$ & 14.00$\pm$0.03 mag & 1 \\
$H$ & 13.28$\pm$0.04 mag & 1 \\
$K_s$ & 12.83$\pm$0.03 mag & 1 \\
$I-J$ & 3.80$\pm$0.15 mag & 1,7 \\
$J-H$ & 0.72$\pm$0.05 mag & 1 \\
$H-K_s$ & 0.45$\pm$0.05 mag & 1 \\
$J-K_s$ & 1.17$\pm$0.04 mag & 1 \\
\enddata
\tablenotetext{a}{Equinox J2000 coordinates at epoch 10 May 1999 from 2MASS.}
\tablenotetext{b}{Estimated; see $\S$~3.4.}
\tablenotetext{c}{Assuming this source is single; see $\S$~4.2.1.}
\tablenotetext{d}{Based on the absence of {\lii} absorption at 6708~{\AA}; see $\S$~3.1.}
\tablerefs{(1) 2MASS \citep{skr06}; (2) SSS \citep{ham01a,ham01b,ham01c}; (3) This paper; (4) \citet{gol04}; (5) \citet{vrb04}; (6) \citet{bur97};
(7) DENIS \citep{epc97}.}
\end{deluxetable}

\begin{deluxetable}{llllllll}
\tabletypesize{\footnotesize}
\tablecaption{Near-infrared Classifications of Blue L dwarfs in Figure~\ref{fig_blue}. \label{tab_classblue}}
\tablewidth{0pt}
\tablehead{
 & \colhead{Optical} & & & & & & \colhead{NIR} \\
 \colhead{Source} &
\colhead{SpT} &
\colhead{{\wat}-A\tablenotemark{a}} &
\colhead{{\wat}-B\tablenotemark{a}} &
\colhead{K1\tablenotemark{a}} &
\colhead{{\wat}-H\tablenotemark{b}} &
\colhead{{\wat}(c)\tablenotemark{b,c}} &
\colhead{SpT\tablenotemark{d}}  \\
}
\startdata
2MASSW~J1300425+191235 & L1 & 0.62 (L3.5) & 0.69 (L3.5) & 0.24 (L2.5) & 0.76 (L4.5) & 1.04 (L0.5) & L3$\pm$0.5 \\
SIPS~J0921-2104 & L2 & 0.56 (L5.5) & 0.67 (L4) & 0.25 (L2.5) & 0.74 (L5.5) & 0.95 (L2) & L4$\pm$1.5 \\
2MASSI~J1721039+334415 & L3 & 0.54 (L6) & 0.62 (L5.5) & 0.30 (L4) & 0.71 (L7) & 0.90 (L3) & L5$\pm$1  \\
{\name} & L4.5 & 0.46 (L8.5) & 0.57 (L6.5) & 0.34 (L4.5) & 0.65 (L9) & 0.72 (L5.5) & L6.5$\pm$2 \\ 
\enddata
\tablenotetext{a}{Indices and index/spectral type relations from \citet{rei01}.}
\tablenotetext{b}{Indices and index/spectral type relations from \citet{meclass} and this paper (Eqn.~3 and 4).}
\tablenotetext{c}{Color-corrected {\wat} index.}
\tablenotetext{d}{Based only on the \citet{rei01} indices.}
\end{deluxetable}

\begin{deluxetable}{lcccccl}
\tabletypesize{\footnotesize}
\tablecaption{Blue L Dwarfs Reported in the Literature. \label{tab_blue}}
\tablewidth{0pt}
\tablehead{
\colhead{Source} &
\colhead{Spectral Type} &
\colhead{$V_{tan}$} &
\colhead{$J$} &
\colhead{$J-K_s$} &
\colhead{$\Delta(J-K_s)$\tablenotemark{a}} &
\colhead{Ref.} \\
 & 
\colhead{Optical/NIR} &
\colhead{({\kms})} &
\colhead{(mag)} &
\colhead{(mag)} &
\colhead{(mag)} & \\
}
\startdata
SDSS~J080531.84+481233.0\tablenotemark{b} & L4/L9.5: &  \nodata & 14.73$\pm$0.04 & 1.29$\pm$0.05 & $-0.58$ & 1,2,3,4,5 \\
SIPS~J0921-2104 & L2/L4:\tablenotemark{c} &  58 & 12.78$\pm$0.02 & 1.09$\pm$0.03 & $-0.62$ & 6,7,8,9 \\
SDSS~J093109.56+032732.5 & --/L7.5: &  \nodata & 16.62$\pm$0.14 & $<$0.88 & \nodata & 2 \\ 
SDSS~J103321.92+400549.5 & --/L6 &  \nodata & 16.64$\pm$0.16 & $<$1.24  & \nodata & 3 \\ 
SDSS~J112118.57+433246.5 & --/L7.5 &  \nodata & 17.01$\pm$0.20 & 1.49$\pm$0.29  & \nodata & 3 \\ 
{\name} & L4.5/L6.5:\tablenotemark{c} & 117 & 14.00$\pm$0.03 & 1.17$\pm$0.04 & $-0.82$ & 8,10 \\
SDSS~J142227.25+221557.1 & --/L6.5: &  \nodata & 17.06$\pm$0.18 & 1.42$\pm$0.25  & \nodata & 3 \\ 
2MASSW~J1300425+191235 & L1/L3.5\tablenotemark{c} & 98 & 12.72$\pm$0.02 & 1.09$\pm$0.03 & $-0.33$ & 9,11,12,13 \\
SDSS~J133148.92-011651.4 & L6/L8: &  \nodata & 15.46$\pm$0.04 & 1.39$\pm$0.08 & $-0.34$ & 1,2 \\
2MASSI~J1721039+334415 & L3/L5:\tablenotemark{c} & 139 & 13.63$\pm$0.02 & 1.14$\pm$0.03 & $-0.58$ &  9,12,13 \\
\enddata
\tablenotetext{a}{Difference in $J-K_s$ color from the average of L dwarfs
with similar optical spectral type \citep{kir00}.}
\tablenotetext{b}{This source appears to be an unresolved binary system
\citep{me0805}.}
 \tablenotetext{c}{See Table~\ref{tab_classblue}.}
\tablerefs{(1) \citet{haw02}; (2) \citet{kna04}; (3) \citet{chi06}; (4) \citet{me0805}; (5) J.\ D.\ Kirkpatrick et al.\ (in preparation); (6) \citet{dea05};
(7) I.\ N.\ Reid et al.\ (in preparation); (8) This paper; (9) \citet{sch07};  
(10) \citet{fol07}; (11) \citet{giz00}; (12) \citet{cru03}; (13) \citet{cru07}}
\end{deluxetable}

\clearpage

\clearpage

\clearpage

\clearpage

\clearpage

\clearpage

\clearpage

\clearpage

\clearpage

\clearpage

\clearpage

\end{document}